\documentclass[aip,amsmath,amssymb,reprint]{revtex4-1}

\usepackage{graphicx}
\usepackage{dcolumn}
\usepackage{bm}

\usepackage[utf8]{inputenc}
\usepackage[T1]{fontenc}
\usepackage{mathptmx}

\newcommand{\beq}{\begin{equation}}
\newcommand{\eeq}{\end{equation}}
\newcommand{\bea}{\begin{eqnarray}}
\newcommand{\eea}{\end{eqnarray}}

\def\eq#1{Eq.~(\ref{#1})}
\def\fig#1{Fig.~\ref{#1}}
\def\app#1{Appendix~\ref{#1}}
\def\refc#1{Ref.~\cite{#1}}

\usepackage[dvipsnames]{xcolor}

\begin{document}

\title{Domain formation in bicomponent vesicles induced by composition-curvature coupling}

\author{Julie Cornet}
\email{cornet@irsamc.ups-tlse.fr}
\author{Nicolas Destainville}
\email{nicolas.destainville@univ-tlse3.fr}
\author{Manoel Manghi}
\email{manghi@irsamc.ups-tlse.fr}
\affiliation{ 
Laboratoire de Physique Th\'eorique (IRSAMC), Universit\'e de Toulouse, CNRS, UPS, France.
}

\date{\today}

\begin{abstract}
Lipid vesicles composed of a mixture of two types of lipids are studied by intensive Monte-Carlo numerical simulations. The coupling between the local composition and the membrane shape is induced by two different spontaneous curvatures of the components. We explore the various morphologies of these biphasic vesicles coupled to the observed patterns such as nano-domains or labyrinthine mesophases. The effect of the difference in curvatures, the surface tension and the interaction parameter between components are thoroughly explored. Our numerical results quantitatively agree with previous analytical results obtained by Gueguen \textit{et al.}, Eur. Phys. J. E 2014, \textbf{37}, 76 in the disordered (high temperature) phase. Numerical simulations allow us to explore the full parameter space, especially close to and below the critical temperature, where analytical results are not accessible. Phase diagrams are constructed and domain morphologies are quantitatively studied by computing the structure factor and the domain size distribution. This mechanism likely explains the existence of nano-domains in cell membranes as observed by super-resolution fluorescence microscopy.
\end{abstract}

\maketitle

\section{Introduction}
\label{sec:intro}

The organization of cell membranes at the molecular scale is the subject of intensive research boosted by the recent development of revolutionary super-resolution microscopy techniques~\cite{Lang2010}. One of their most striking features is that their molecular constituants are organized in domains, the size of which ranges from few dozens of nanometers to micrometers, in which few lipid and/or protein species are segregated~\cite{Komura2014,Jacobson2016,Sezgin2017}. Hence cell plasma membranes are now consensually seen as patterned two-dimensional systems, but the physico-chemical mechanisms accounting for these observations are not consensual today~\cite{Leslie2011,Levental2016,Schmid,review}. In particular the reason why membrane domains are as small as dozens of nanometers is still matter of debate.

A common general physical mechanism describes patterning in various soft-matter contexts and for a wide range of length scales~\cite{Seul}: below the demixing temperature, short-range attraction that drives monomer association competes with weaker, however longer-range repulsion that stops the segregation and limits the aggregate size. This mechanism gives rise to modulated phases as we term them in this work.
In the model studied here, assumed to be a binary mixture for sake of simplicity, one of the molecular species, named A, imposes a local spontaneous curvature to the elastic membrane while the other species, named B, does not. In biologically relevant situations, A is assumed to be the minority species. When species aggregate below the demixing temperature, A-domains acquire a curved shape. The membrane being under tension, this leads to the increase of the tension term in the elastic energy. Membrane patterning then results from the energetic competition between two mechanisms~\cite{Leibler,Schick}: on one hand, molecular species interactions  (short-range attraction) leads to phase separation. On the other hand, large A-domains have an elastic cost that can be shown to grow faster than their area. This can eventually make too large domains, and a fortiori macrophases, unstable. This leads to an effective long-range repulsion between A-species molecules and to the formation of smaller structures in equilibrium~\cite{review}. This mechanism explains the formation of meso-domains or labyrinthine structures (stripes), depending on A-species concentration. An additional interest of such a mechanism is that it remains efficient {\em above} the demixing temperature, where membrane shape fluctuations stabilise structured composition fluctuations (see Ref.~\cite{GG1} and references therein).

In this Article, we study a numerical model of bicomponent, tessellated (i.e. triangulated) membrane, described by two fields: the membrane height field, ruled by elastic free energy, and the composition field, ruled by mixture free energy. These two fields are coupled since one of the species imposes a local spontaneous curvature to the membrane.
By contrast, we suppose for sake of simplicity that the bending modulus $\kappa_0$ is uniform on the whole membrane, whereas some models assume that it depends on the composition~\cite{review}. This numerical approach extends to the spherical geometry the planar model of Ref.~\cite{Wallace2005}. We choose $\kappa_0=20~k_{\rm B}T$ (where $k_{\rm B}T\simeq 4\times 10^{-21}$~J is the thermal energy at room temperature), a typical value for biomembrane lipids~\cite{Mouritsen,Phillips}. The situation where $\kappa_0$ also depends on $\phi$ will be studied in a forthcoming article. One of our goals is to propose a numerical verification of the analytical studies provided for example in Refs.~\cite{Leibler,Hansen1998,Kumar1999,Schick,GG1}, which relied upon some approximations. In particular, the Gaussian or mean-field theories studied there were not expected to be valid below or close to the mixture critical temperature, and we also intend to address the system properties in this case of potential biophysical interest. It is possible to derive some exact analytical solutions below the critical temperature~\cite{Jiang2000}, however, this requires to neglect thermal fluctuations and to assume restrictive symmetries.

We expect to find numerically the phase diagrams predicted by analytical calculations, displaying four characteristic phases: macro-, disordered (or dilute),  ``structured disordered'' (or microemulsion), and ``structured ordered'' (or meso-) phases. The two last ones are the modulated phases that we have just mentioned, above and below the demixing temperature. In addition, analytical predictions are based on the calculation of structure factors that give access to typical wavelengths of modulated phases. However, they do not provide any information on the shape of patterns that can be roundish domains, elongated ones, stripes or even more complex, labyrinthine morphologies. Our numerical simulations can give access to such information. 

In addition, it is our ambition to propose a model able to explain the experimentally observed size of nano-domains on the surface of eukaryotic cells, below the diffraction limit. At a \emph{qualitative} level, in Ref.~\cite{review}, we have argued that a model based upon the competition between attraction at short range and weak repulsion at longer range likely explains the existence of such nano-domains.  Using realistic values of the numerical parameters entering the model, this work will demonstrate that such an approach remains realistic at a \emph{quantitative} level, in vesicle geometry. To our knowledge, this has never been achieved so far in this context.

One important particularity of our numerical model is that, contrary to alternative models~\cite{Lipowsky, Amazon, Penic}, the area is not locally constrained at the scale of elementary triangles but globally controlled by surface tension and imposed volume~\cite{GG2}. Consequently, the imposed surface tension is known exactly and is not affected by local constraints in an ill-controlled manner. Moreover imposing a surface tension is more realistic in modeling a real vesicle, the area of which fluctuates.

It is also worth mentioning that at the level of coarse-graining of the model, where an elementary membrane patch represents many molecules, the model embrasses several experimental situations: A-species can either represent a different lipid phase, e.g. liquid-ordered domains in an otherwise liquid-disordered sea~\cite{Sezgin2017,Hossein2020}, in which some attracting curving molecules can also be incorporated~\cite{Shimobayashi}; or an otherwise homogeneous lipid phase locally enriched in some curving proteins. Indeed, spontaneous curvature can arise from either different lipid composition between both membrane leaflets, or proteins inserted in the membrane, breaking in both cases the up/down symmetry~\cite{Mouritsen,review}. 

The paper is organized as follows. After presenting the mesoscopic model in Section~\ref{sec:model} and its numerical implementation in Section~\ref{sec:num_meth}, we present our results in the main Section~\ref{results}. Our main findings are a quantitative description of the effect of both curvature coupling and surface tension on the domain morphology connected to the vesicle deformation. This is synthesized in phase diagrams for different global compositions, constructed by using the numerical structure factors. These phase diagrams display the different expected phases in agreement with previous analytical results above the demixing temperature. Below this temperature, the domain size distributions are characterized and compared to experimental observations.

\section{Mesoscopic model}
\label{sec:model}

A model biomembrane can be seen as a bidimensional fluid mosaic, described by the elastic free energy of its surface and the interaction of its components. This work develops further a previous study by Gueguen \textit{et al.}~\cite{GG1}.

The Canham-Helfrich elastic free energy of the membrane is~\cite{Helfrich}
\bea
H_{\rm Helf} =\frac{1}{2}\int_\mathcal{A}  \kappa (H - C)^2  d S + \sigma \mathcal{A} 
\label{eq:Helf}
\eea
where $\mathcal{A}$ is the membrane area.
The first term of the rhs. of \eq{eq:Helf} is the elastic contribution, with the bending modulus $\kappa$, constraining locally the membrane to its spontaneous curvature $C$, where $H$ is the local curvature. In the case of a closed vesicle of interest in this work,  the membrane is described by a height function $u(\theta,\varphi)$ measuring the distance to a reference sphere of radius $R$ with ${\bf r}(\theta,\varphi)=R[1+u(\theta,\varphi)] {\bf e}_r$  the position of the membrane. When membrane fluctuations remain small, $H \simeq \frac2{R} [1- u-\frac12 (\partial^2_\theta u+\frac1{\sin^2\theta}\partial^2_\varphi u) +\mathcal{O}(u^2)]$ (see for instance~\cite{GG1}). Note that curvature is assumed to be positive when the membrane is convex.
The bending modulus $\kappa$ typically falls  in the 10 to $100\ k_{\rm B}T$ interval for biomembranes \cite{Mouritsen, Dimova}.
The surface tension is denoted by $\sigma$ which appears as a Lagrange multiplier controlling the membrane area. There exist several alternative definitions of the surface tension, coinciding in the high tension limit~\cite{GG2}. Here we consider values of $\sigma$ on the order of $10^{-8}$~J.m$^{-2}$ allowing moderate shape fluctuations~\cite{Bassereau}. The vesicle is then considered to have a quasi-spherical shape fluctuating around the sphere of radius $R$. 

In addition, we consider a membrane made of a binary mixture of two species, A and B, and the corresponding composition fields are $\phi_A$ and $\phi_B$. We note $\phi=\phi_A$ (thus $\phi_B=1-\phi$). The interaction between membrane components of the mixture is described by a Ginzburg-Landau Hamiltonian~\cite{Chaikin}:
\begin{equation}
H_{\rm GL}[\phi] = \int_\mathcal{A} {\rm d}S \left[ \frac{m}2 (\phi-\phi_c)^2 + \frac{b}2 (\nabla \phi)^2 \right]
\label{eq:GL}
\end{equation}
where $\phi(\mathbf{r}) \in [0,1]$ is the composition field, and $\phi_c =1/2$ is the critical composition. The mixture undergoes a phase transition at a critical temperature $T_c$. The potential $\frac{m}2 (\phi-\phi_c)^2 $ ensures a homogeneous phase for high enough temperatures $T>T_c$ ($m>0$) and a phase separation for lower temperatures $T<T_c$ ($m<0$). Close to $T_c$, $m\propto k_{\rm B}(T-T_c)/a^{2}$ where $a$ is a microscopic (UV) cutoff. In the case of phase separation, the mixture gives two distinct phases, one rich in A and the other one rich in B. One can then define the line tension $\lambda$ corresponding to the energy cost of the interface per unit length. Close to $T_c$, $\lambda=\lambda_0(T_c-T)$ for the 2D Ising universality class~\cite{Chaikin} where $\lambda_0$ is a constant. The term $\frac{b}2 (\nabla \phi)^2$ characterizes the energy cost ensuing from the local variation of composition, where $b$ is the so-called stiffness. Below $T_c$ positive terms in $\phi^4$ need to be considered to enable the partition function to converge that would diverge otherwise.

From a biophysical perspective, we consider that the locally higher curvature of the membrane can either be induced by lipid mixture symmetry-breaking~\cite{Mouritsen,Hossein2020} or be imposed by integral or peripheral proteins~\cite{review}. In both cases, $\phi(\mathbf{r})$ then measures the local density of the curving molecules, lipids or proteins, that we call A-species in this work. 

To introduce this coupling between the composition and the membrane curvature, it is usually considered that the local spontaneous curvature and/or the bending modulus are functions of the local concentration, $C(\phi)$ and $\kappa(\phi)$. As a first approximation, one can choose a linear form of the couplings, as in \refc{GG1}: 
\begin{align}
C=C_0+C_1\phi \\
\kappa=\kappa_0+\kappa_1\phi
\end{align}
In the present work we assume that the bending modulus is not dependent on the phase, i.e. $\kappa_1=0$. The term $C_1$ is the difference between the spontaneous curvatures of the two phases, pure A ($C_0+C_1$) and pure B ($C_0$). Hence in Eq.~(\ref{eq:Helf}) we include a term accounting for the coupling:
\begin{align}
\frac{\kappa_0}2 (H - C)^2=\frac{\kappa_0}2 [H - (C_0+C_1\phi)]^2
\label{eq:coupling}
\end{align}
One can define the coupling strength $\Lambda=-\kappa_0 C_1$ between the fields $H$ and $\phi$ (see Ref.~\cite{review}). In our particular case, we always consider the spontaneous curvature of the majority B-species as being the spontaneous curvature of the sphere of radius $R$, i.e. $C_0=2/R$.

The analytical study of the lipid binary mixture was carried out in \refc{GG1}. Our case in the current numerical work corresponds to this description with only one composition field $\phi$ ($\phi-\phi_c$ is noted $\psi_-$ in \refc{GG1}).
In order to study the influence of the different parameters on the formation of modulated phases, we will construct a phase diagram. To do so, we first study the structure factor of the system that provides information about its degree of structure~-- and also characterizes the amplitude of the response of the local composition to an external perturbation. Indeed, when a system features domains, it holds underlying order, i.e. modulated density fluctuations. 

\begin{table}[h]
\small
\caption{Main dimensionless parameters used in this article ($R$ is the vesicle radius)}
\label{tab}
  \begin{tabular*}{0.48\textwidth}{@{\extracolsep{\fill}}lll}
\hline
Parameter & Expression & Defined in \\
\hline
Spontaneous curvature & $c_0=C_0R$;  $c_1=C_1R$ & Section~\ref{sec:model} \\ 
Bending modulus & $\tilde{\kappa}_0=\kappa_0/k_{\rm B}T$ & Section~\ref{sec:model} \\ 
Surface tension & $\tilde{\sigma}=\sigma R^2/k_{\rm B}T$; $\hat{\sigma}=\sigma R^2/\kappa_0$& Section~\ref{sec:model} \\
Ising parameter & $\tilde{J}_I=J_I/k_{\rm B}T$ & Section~\ref{sec:model}  \\
 & $\hat{J}=2\sqrt{3}J_I/\kappa_0$ & \app{sec:link} \\
Ginzburg-Landau  & $\hat{m}=m R^2/\kappa_0$  & \app{sec:link}\\
 parameter (mass)    & $=\alpha_0 N(1-J_I/J_{I,c})/\tilde{\kappa}_0$  & \app{sec:link} \\
\hline
\end{tabular*}
\end{table}

The fixed concentration in A-species in the vesicle is 
\beq
\bar{\phi}= \frac{1}{\mathcal{A}}\int_\mathcal{A} \phi(\mathbf{r}) {\rm d}S
\eeq
By rotational symmetry, the angular correlation function of the composition field fluctuation $\psi=\phi-\bar\phi$ is a function of the angle between any two points on the vesicle. It writes (see Ref.~\cite{GG1} for further details)
\beq
\langle \psi(\theta) \psi(0) \rangle = \frac{k_{\rm B}T}{4 \pi \kappa_0} \sum_{l \geq 1} \frac{(2l+1) P_l(\cos \theta)}{M(l)}
\label{eq:correl}
\eeq
where the $P_l$ are Legendre polynomials~\cite{Abramowitz}. Then one shows that (see \app{SF})
\beq
M(l)=\hat{m}+ \frac{c_1^2 \hat{\sigma}}{l(l+1)-2+\hat{\sigma}} + 2\hat{J}l(l+1)
\label{eq:M}
\eeq
where dimensionless parameters are introduced and presented in table~\ref{tab}, the lengths being divided by the radius $R$ and the energies by $k_{\rm B}T$ or $\kappa_0$.

The structure factor is defined as the coefficients of the Legendre polynomials in \eq{eq:correl}:
\beq
S(l)= \int^\pi_0 \langle \psi(\theta) \psi(0) \rangle P_l(\cos \theta) \sin \theta d \theta
\label{Scalc}
\eeq
which yields
\beq
S(l)=\frac{k_{\rm B}T}{2\pi \kappa_0}\frac{1}{M(l)}
\label{eq:SF}
\eeq
In the following, we fit the structure factors obtained numerically with this expression.

\section{Numerical implementation of the model}
\label{sec:num_meth}

In our numerical simulations we consider the vesicle as a tessellated sphere composed of $N$ vertices~\cite{GG2}. On each vertex stands a patch of one of the two species. The size of this patch, both in terms of diameter and number of molecules, is tunable and depends on the vesicle radius $R$ (set to 1 in the simulations). We can get different system sizes $N$ by choosing the number of times that we iterate the subdivision process in the sphere tessellation (see \app{sec:tess} and Ref.~\cite{GG2}). As an example, for a simulation with $N=2562$ sites, in a vesicle of radius $10~\mu$m, a patch contains $\sim 10^6$ lipids.

The elastic free energy in \eq{eq:Helf} is discretized as follows, with the help of the Laplace-Beltrami operator for the curvature term:
\beq
H_{\rm Helf}= \frac{1}{2}\sum_i \kappa_0 [2H_i - C_i]^2 \mathcal{A}_i + \sigma \sum_i \mathcal{A}_i
\eeq
with $\mathcal{A}_i$ the area associated to a vertex. The term $2H_i$ is the norm of the Laplace-Beltrami operator $\mathbf{K}_i$ (total curvature) computed following Ref.~\cite{Meyer} as
\beq
\mathbf{K}_i=\frac{1}{2\mathcal{A}_i}\sum_j(\cot \alpha_{ij} + \cot \beta_{ij})(\mathbf{x}_i-\mathbf{x}_j)
\eeq
where $\mathbf{x}_i$ is the position of vertex $i$ and the sum is taken over the neighbors of $i$. The angles $\alpha_{ij}$ and $\beta_{ij}$ are the angles of the two triangles sharing the edge $\mathbf{x}_i \mathbf{x}_j$ and opposite to this edge. See \refc{GG2} and Fig. 6 therein for illustration. The uniform bending modulus is set to $\kappa_0=20~k_{\rm B}T$. In our simulations, the vesicle volume is fixed close to the volume of the initial sphere $V_0$ by a hard quadratic constraint. By contrast the total vesicle area is constrained by a soft constraint and controlled by the surface tension, acting as a Lagrange multiplier and allowing the surface to fluctuate reasonably:
\beq
H_{\rm c}= \sigma \mathcal{A} + \frac{1}{2}K_v\left(\frac{V}{V_0}-1\right)^2
\eeq
where $K_v=2\times10^6 k_{\rm B}T$. Contrary to other studies~\cite{Lipowsky, Amazon, Penic} we impose a global constraint on the total vesicle area and do not introduce local constraints on the triangle edge lengths. These local bounds induce resulting forces on the edges, thus influencing the surface tension and making its value difficult to control while it plays a crucial role in membrane spatial organization \cite{GG2}. Since we are here interested in weak shape deformations and considering the vesicle in equilibrium, we are not concerned about dynamical aspects and thus allowing edge flipping in the tesselated system is not required. In addition, since in our model a site is not assigned to a specific patch of lipids, the lipid patches diffuse freely on the lattice and we do not need to apply edge flipping moves to account for membrane fluidity. This turns needful if one wishes to study membrane dynamics and large deformations as the one at play in phenomena such as crumpling~\cite{gompper_crumpling_1995}.

The discrete two-dimensional Ising (or lattice-gas) model, very relevant to study phase transition phenomena, is used to describe the binary mixture. It belongs to the same universality class as the Ginzburg-Landau continuous theory in \eq{eq:GL} and these two models are equivalent above $T_c$ in the continuous limit \cite{Chaikin}. The discrete Hamiltonian between nearest-neighbor sites on the lattice is:
\beq
H_{\rm Ising}= -J_I\sum_{\langle i,j \rangle} s_i s_j 
\eeq
where $s_i = \pm 1$. The composition on any vertex $i$ of the tessellated lattice is $\phi_i =0$ or 1, related to $s_i$ through $\phi_i =(1+s_i)/2 $.
The Ising parameter $J_I>0$ measures the tendency of the species to demix. It is related to the parameter $b$ of \eq{eq:GL} via $b = 4 \sqrt{3} J_I$ on a triangular lattice (see \app{sec:link}). Note that only the first neighbors come into play, mimicking short range (e.g., van der Waals) interactions between membrane constituents.

Varying the temperature $T$ in the simulations of a pure Ising model amounts to tuning the species affinity via the interaction parameter $J_I$. In our simulations we rather fix the temperature $T$ at room temperature and we vary the value of $J_I$. In this way, the temperature of the fluctuating membrane is kept fixed.  For $J_I<J_{I,c}$ (respectively $J_I>J_{I,c}$) we have a disordered (resp. ordered) phase with the critical value $J_{I,c}\simeq k_{\rm B}T/3.64$ on an infinite triangular lattice~\cite{Baxter}. 
\footnote{We have measured numerically the critical value of $J_I$ by computing the specific heat at $\bar \phi =0.5$ without curvature coupling (pure Ising model). Consistently, it has a maximum at $\tilde{J}_{I,c}^{-1}\simeq3.62$. Of course it is different from the value found with mean-field $\tilde{J}_{I,c}^{-1}=6$, the number of first neighbors in a triangular lattice.}

One can relate $m$ and $J_I$ through $\hat{m}=\frac{\alpha_0 N}{\tilde{\kappa}_0}(1-\frac{J_I}{J_{I,c}})$ with $\alpha_0>0$ (see \app{sec:link}). A mean-field approximation for $\alpha_0$ can be drawn from Flory theory~\cite{Flory}, $\alpha_0=1/\pi$. We introduce the dimensionless Ising parameter varying in our simulations $\tilde{J}_I=J_I/k_{\rm B}T$ and then $\tilde{J}_{I,c}^{-1}\simeq3.64$.

\begin{figure}[t]
\centering
\includegraphics[width=0.9\columnwidth]{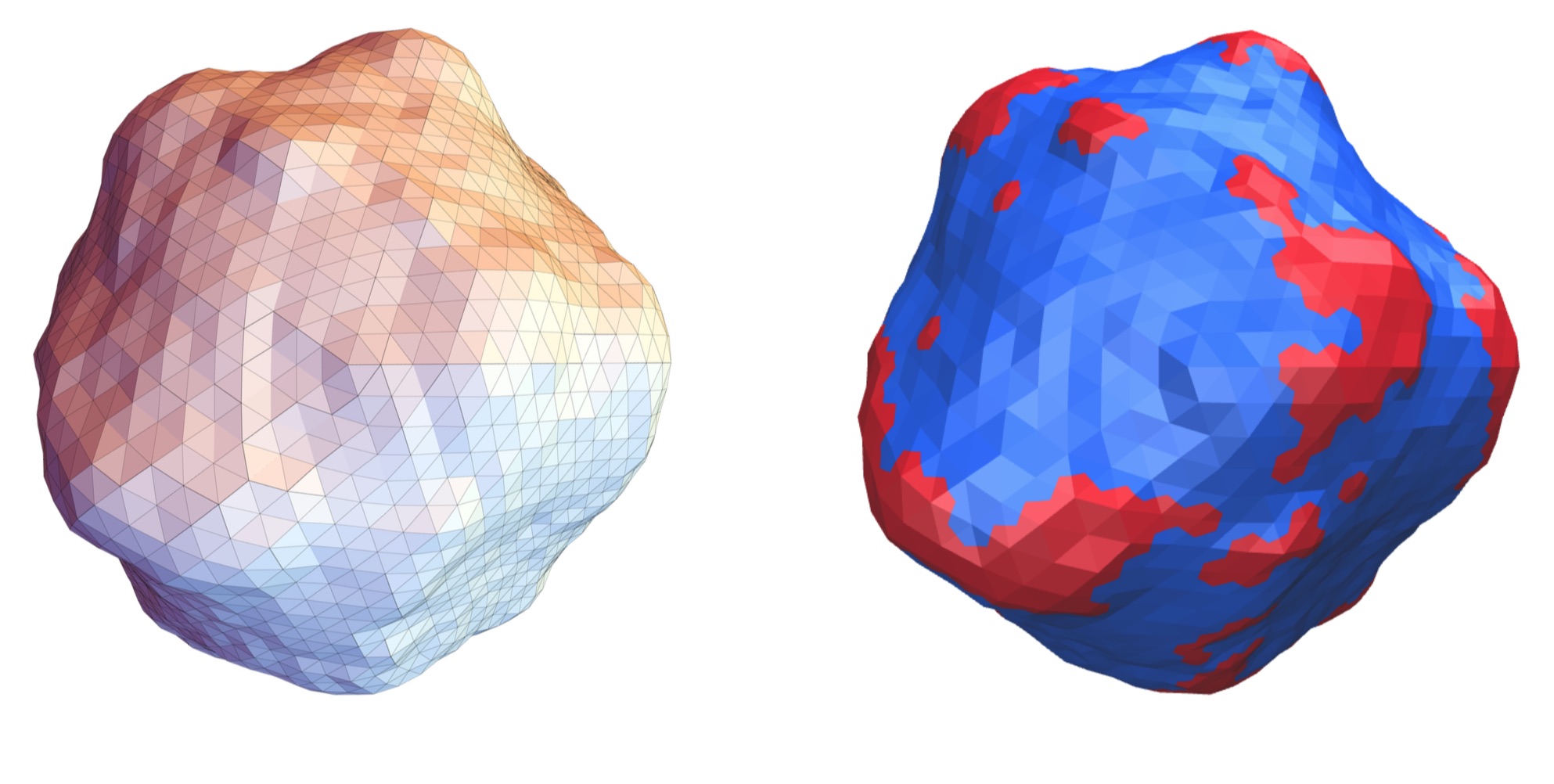} 
\caption{Snapshot of a tessellated vesicle in equilibrium using the Metropolis/Kawasaki algorithm, showing the membrane height field $u$ (left) and the composition one $\phi$ (right), A (resp. B) species in red (resp. blue).}
\label{fig:num_method}
\end{figure}
 
We have implemented a Monte Carlo (Metropolis) algorithm. More precisely, since a system with conserved order parameter $\bar \phi$ is considered, we use the Kawasaki algorithm~\cite{Newman} for the composition field. At each step of the program, two local moves are applied to the vesicle on random vertices that are accepted or not: (1)~a vertex undergoes a small radial displacement, which locally modifies the elastic energy; (2)~the spin values of two neighbor vertices are swapped following the Kawasaki prescription, modifying the interaction energy.  The iteration of this process converges to the equilibrated three dimensional conformations of the membrane (shape and component spatial organization, see \fig{fig:num_method}). We measured correlation times to be typically $10^8$ Monte Carlo (MC) steps for both shape and composition fields for a system of $N=2562$ sites. For high $\tilde{J}_I$ (low Ising temperatures) and low coupling $c_1$, the dynamics is slow. Indeed, once a macrophase is formed, the macrocluster exchanges elements by Ostwald ripening process only and diffuses slowly~\cite{Chaikin}. However, we are not interested in describing the phenomenon that happens at these large time scales but rather in the equilibration of cluster size distributions and structure factors as discussed below. We thus performed simulations of $10^{10}$ MC steps each so that we have good sampling for the different measured observables, averaged on $\sim 10^3$ independent configurations. We also performed some simulations with $N=10242$ vertices. However, the excessive simulation time required to obtain good sampling for this system size restrained us to a few parameter sets only and lower statistical sampling. All the systems studied in this work are in thermodynamical equilibrium. 

Triangulating the sphere leads to the construction of triangles of slightly different surfaces (the largest triangles are typically 10\% larger than the smallest ones). In our numerical model, the bending energy of a vertex is proportional to the area associated with it~\cite{GG2}. Thus the most curved regions tend to get localized to the smallest triangles, close to the 12 vertices of coordination number 5, which biases the free energy minimisation. We corrected this issue and reduced the main effect by decreasing the triangle area dispersion by a factor $\sim 100$. However, a slight bias still persists. See \app{sec:tess} for more detailed explanations.

To compare numerical results to available analytical predictions and experimental data, one regularly computes different observables throughout the simulation, once the system has reached equilibrium. In particular, we measure temporal and spatial correlation functions for the height function $u$ and the composition field $\phi$.

From the composition field correlation function $\langle \psi(\theta) \psi(0) \rangle$, we compute the structure factor $S(l)$ following \eq{Scalc}. Note that the physical maximum value $l_{\rm max}$ of $l$, related to the UV cut-off satisfies $(l_{\rm max}+1)^2=N$ in order to have the same number of degrees of freedom in both direct and reciprocal spaces~\cite{GG1}. For $N=2562$ it gives $l_{\rm max}=49$. In practice the integral \eq{Scalc} is discretized because the correlation function $\langle \psi(\theta) \psi(0) \rangle$ is measured as a histogram.

\section{Results}
\label{results}

\subsection{Curvature-composition coupling effect on domain formation}
\label{sec:effetC}

We performed Monte Carlo simulations of vesicles for various sets of parameters, varying the curvature coupling strength $c_1$, the surface tension $\tilde{\sigma}$ and the component interaction parameter $\tilde{J}_I$.
The theory developed in \refc{GG1} predicts four different phases arising from the combination of these parameters.
At low curvature coupling $c_1$, the systems are either phase-separated for high $\tilde{J}_I$, or disordered for low $\tilde{J}_I$, reproducing the expected behavior without coupling. At high enough curvature coupling, A-rich domains of various sizes appear, i.e. stable modulated phases. 

\begin{figure}[t]
\centering
\includegraphics[width=0.8\columnwidth]{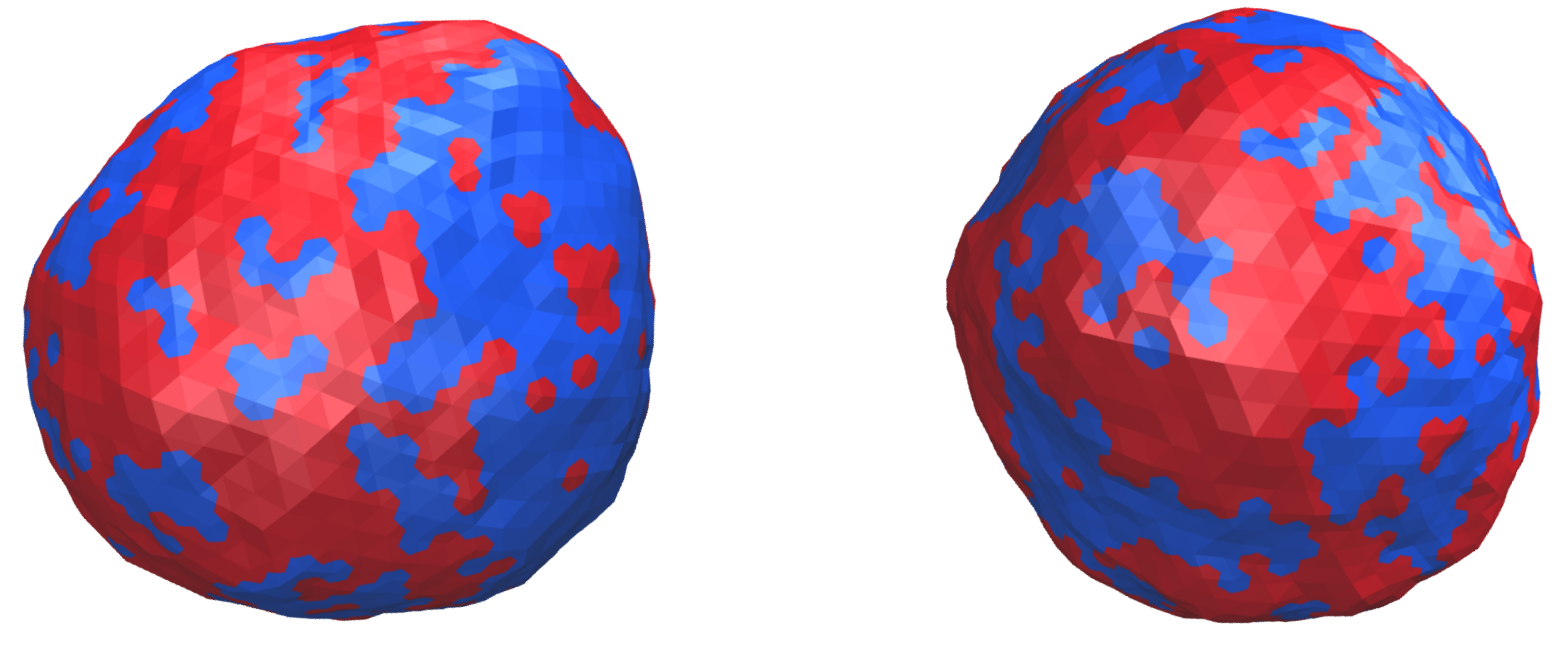} 
\caption{Snapshots of simulated vesicles showing the effect of the difference of spontaneous curvature $c_1=c_2-c_0$ between the two species at low $\tilde{J}_I$ ($\tilde{J}_I^{-1}=4.0$, $\bar{\phi}=0.5$, $\tilde{\sigma}=300$): $c_1=0$ (left) and $c_1=3$ (right).  For the blue species, $c_0=2$ in all the simulations. The two species tend to mix but the coupling $c_1$ stabilises modulated phases, although difficult to catch with the eye (see text).}
\label{fig:lowJ}
\end{figure}

At low $\tilde{J}_I<\tilde{J}_{I,c}$, the two species tend to mix as shown in \fig{fig:lowJ}, but we observe that strong enough coupling of the composition field to shape fluctuations stabilises more ordered local composition fluctuations. Although hardly detectable with the eye, this underlying order is present and can be expressly detected thanks to the computation of the structure factor of the system as we will describe it further in \ref{sec:SF} (see also \fig{fig:SF_diffC_aboveTc}).
We will see that our numerical results are in quantitative agreement with previous analytical studies~\cite{GG1,Schick,Shlomo} (see also the Review article~\cite{review}). 

At large $\tilde{J}_I>\tilde{J}_{I,c}$, the previous analytical description is not adapted anymore as the functional integrals are no more Gaussian because terms in $\phi^4$ must be kept in \eq{eq:GL}~\cite{GG1}. Here comes the interest of the numerical study that can extrapolate the model to these cases.

\begin{figure*}[t]
\centering
\includegraphics[width=1.6\columnwidth]{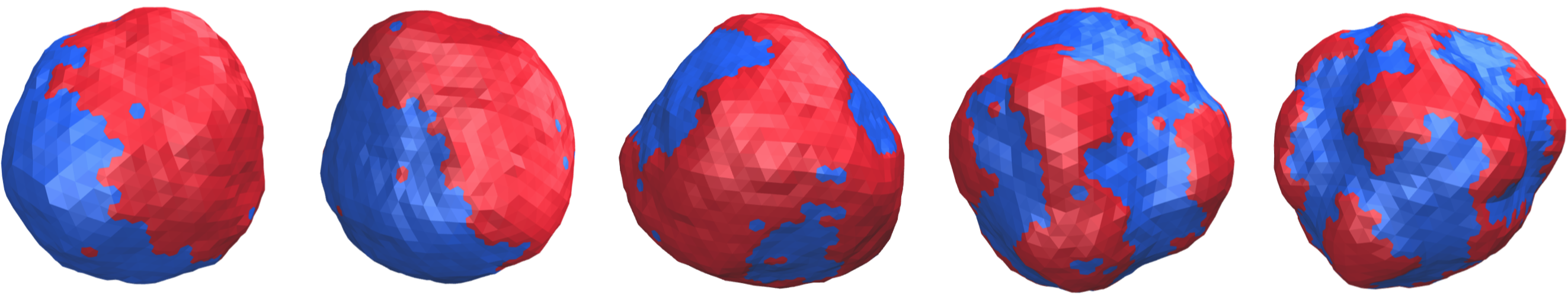} 
\caption{Same as \fig{fig:lowJ} at high $\tilde{J}_I$ with $c_1=0,2,3,4,6$ from left to right. Other parameter values are $\tilde{J}_I^{-1}=2.5$, $\bar{\phi}=0.5$, $\tilde{\sigma}=300$.}
\label{fig:effetC05}
\end{figure*}

Without any curvature coupling, the vesicle undergoes phase separation as expected (Fig.~\ref{fig:effetC05}). When we increase the coupling $c_1$ (from left to right), the large curved domains become unstable and break into smaller ones, getting smaller as  $c_1$ increases.
We see that when we couple the concentration field to shape fluctuations, a system in the macrophase regime $\tilde{J}_I>\tilde{J}_{I,c}$ can move over the phase transition and feature domains. We get modulated phases as shown in \fig{fig:effetC05}. We can then consider an effective Ising parameter for the coupled system modified by the curvature coupling that introduces a new term in $\phi^2$ in \eq{eq:coupling}. It shifts the mass $m$ to:
\beq
m'=m+\kappa_0 C_1^2
\eeq
The transition in a coupled system now occurs when $m'=0$, that is to say when $m=-\kappa_0 C_1^2$. One can then express the effective transition Ising parameter for a coupled system as 
\beq
J_{I,c}^{\rm eff}=J_{I,c}\left(1+\frac{\tilde\kappa_0 c_1^2}{\alpha_0 N}\right)
\label{eq:Jceff}
\eeq
The coupling $c_1$ increases the effective value of the transition Ising parameter as found in \fig{fig:effetC05}, allowing phase modulations even above $\tilde{J}_{I,c}$.
Increasing further $\tilde{J}_I$ drives macrophase separation by increasing line tension. 

To obtain an approximated expression for the domain size, we compute the energy cost due to the area excess $\Delta A$ induced by a curved domain of radius $r$ (we assume spherical cap domains at low enough concentration, as well as spherical vesicle)
\bea
\sigma\Delta A &=& 2\pi\sigma[R_2^2(1-\cos\theta_2)-R_0^2(1-\cos\theta_0)]\nonumber \\
 &\simeq& \frac{\pi}{16}\sigma r^4(C_2^2-C_0^2)
\label{deltaA}
\eea
where $\theta_2$ and $\theta_0$ are the angles of the domain along the osculatory circles of radii $R_2=2C_2^{-1}$ and $R_0=R=2C_0^{-1}$ respectively (related through $r=R_0\sin\theta_0=R_2\sin\theta_2$). The second expression in \eq{deltaA}, obtained by expanding $\Delta A$ at ordre 4 in $r/R_0$ and $r/R_2$, is valid for small domains only, with $r \ll R_2<R$. This energy penalty is balanced by the line energy $2\pi r \lambda$, which yields
\beq
r\simeq2^{5/3} \left(\frac{\lambda}{\sigma}\frac1{C_2^2-C_0^2}\right)^{1/3}
\label{r}
\eeq
Even though ignoring the role of translational and conformational entropies~\cite{Destain2008}, this explain why an increasing $c_1$ (or equivalently $c_2$) favors smaller domains. Since the line tension $\lambda$ is proportional to $J_I-J_{I,c}$~\cite{Veatch}, the higher $J_I$ is, the more difficult it is to form small domains. Note that \eq{r} is very similar to the one obtained by Kawakatsu \textit{et al.} (Eq. (2.12) in~\refc{Andelman1993}) in the strong segregation limit, although obtained with a different argument.

\subsection{Surface tension effect on domain formation}

\begin{figure}[t]
\centering
\includegraphics[width=\columnwidth]{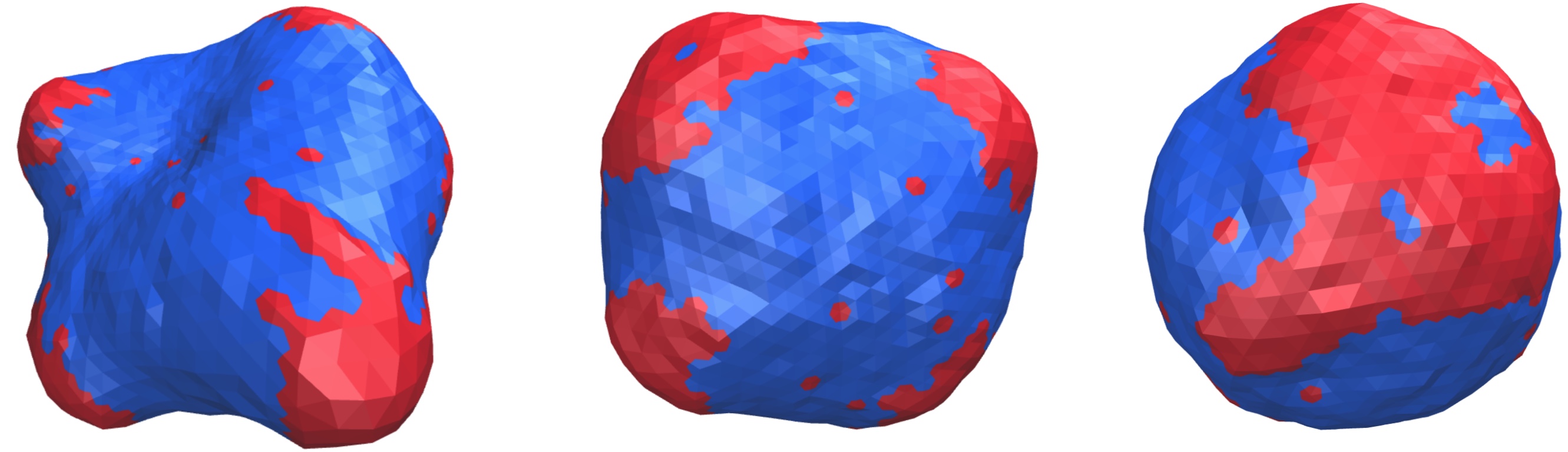} 
\caption{Snapshots of simulated vesicles with increasing surface tension $\tilde{\sigma}$ (respectively from left to right 150, 300 and 600). Other parameter values are $\tilde{J}_I^{-1}=2.5$, $\bar{\phi}=0.2$,  and $c_1=3$.}
\label{fig:effet_sig}
\end{figure}

In \fig{fig:effet_sig}, we show snapshots of simulated vesicles with the same $c_1$ and $\tilde{J}_I$, but with increasing tension $\tilde{\sigma}$. We see that low tensions allow strong membrane deformations. Therefore the formation of domains, induced by curvature coupling, is favored in highly deformed regions. On the contrary, for high surface tensions the vesicle is constrained to a quasi-spherical shape and patterning along with deformation is therefore attenuated or even prevented. 

Thus, at a fixed coupling value $c_1$ leading to mesophases in a low surface tension regime, the system can undergo macrophase separation when the surface tension $\sigma$ is high enough to balance the curvature term $\kappa_0 C_1^2$ in the Helfrich free energy and cancel its effect~\cite{Andelman1992, Andelman1993,review}. Note that \eq{r} is valid for low enough surface tensions such that the domain radius $r$ is smaller than the correlation length $\xi=\sqrt{\kappa/\sigma}$. At higher tensions, the domain shape significantly deviates from a spherical cap of radius $2C_2^{-1}$. Hence \eq{r} applies only to the case of the leftmost snapshot of \fig{fig:effet_sig}, where $\xi \simeq 0.4R$.

Combining the variations of these three key parameters $\tilde{J}_I$, $c_1$ and $\tilde\sigma$, one can get macrophase, disordered or modulated systems that we study in detail below in \ref{sec:phase-diag}. In a certain range of parameters that we will characterize further, we get systems featuring mesopatterning: either mesodomains at low concentration $\bar{\phi}$, or labyrinthine mesophases when $\bar{\phi}$ is high enough so that the A-species percolate through the system (see \fig{fig:effetC05}, right-most vesicles) as predicted for example in Refs. \cite{Kumar1999,Harden2005}, by using a one-mode approximation.

\subsection{Quantitative results and comparison to previous analytical results}

Beyond these qualitative results, we now quantitatively study domain formation thanks to spatial correlation functions and domain size distributions. These observables are computed once the system has reached equilibrium. The aim is to classify the vesicle states and to extract information about the emerging membrane patterns, such as their typical size, spacing or number.

\subsubsection{Structure factor}
\label{sec:SF}

We compute numerical spatial correlation function of the vesicle composition as defined in \eq{eq:correl}. From these measurements we compute the structure factor as described in Sec.~\ref{sec:num_meth}.

\begin{figure*}[t]
\centering
\includegraphics[width=1.9\columnwidth]{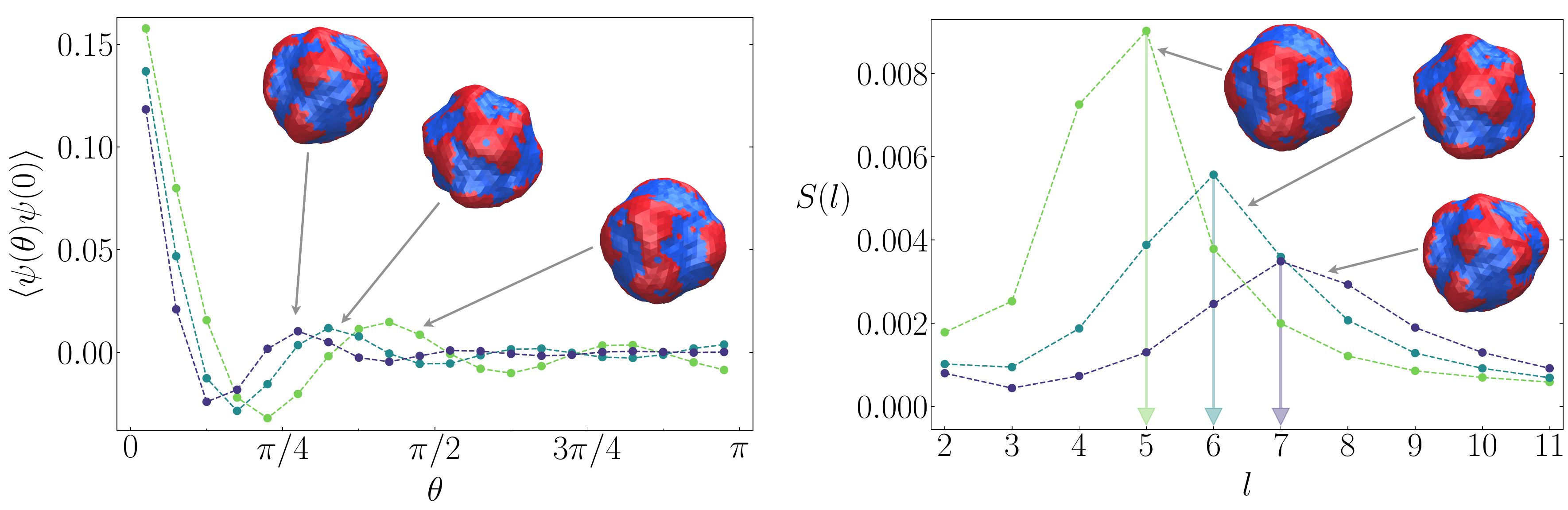} 
\caption{Numerical correlation functions (left) and corresponding structure factors (right) for different curvature coupling values $c_1=4$, 5 and 6 from green to blue ($\tilde{J}_I^{-1}=2.5$, $\bar{\phi}=0.5$, $\tilde{\sigma}=300$). The correlation functions show a first exponential decrease corresponding to pattern width, and secondary oscillations related to pattern spacing. The structure factors exhibit a maximum for an abscissa $l^*$ (arrows), corresponding to pattern wavelength in real space, which increases when $c_1$ increases and the patterns get thinner. Color lines are guides to the eye.}
\label{fig:SF_diffC}
\end{figure*}

\begin{figure}[t]
\centering
\includegraphics[width=0.93\columnwidth]{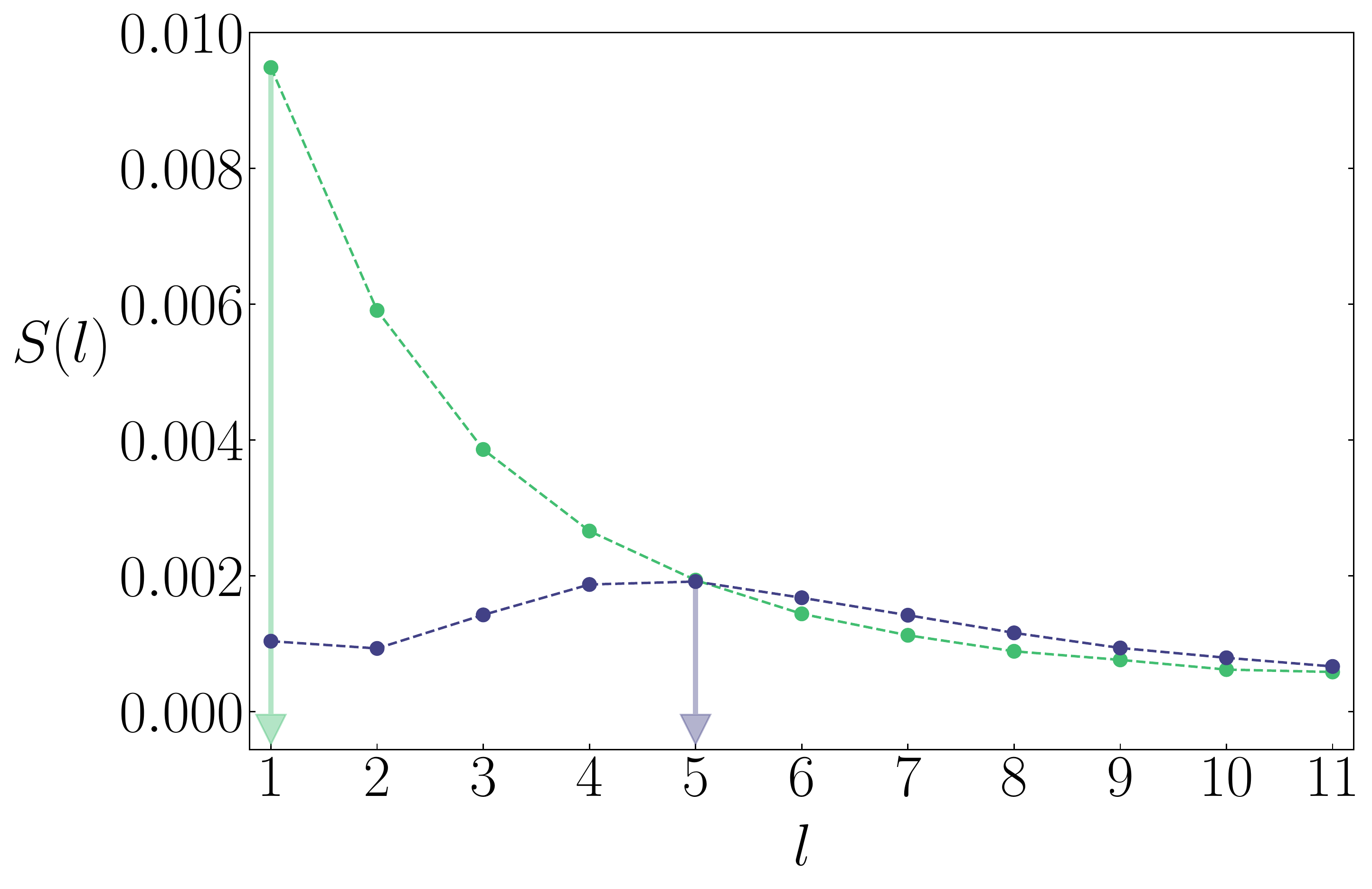} 
\caption{Structure factors for two different curvature coupling values $c_1=0$ (green, no coupling) and $c_1=3.0$ (blue), the ones of the two vesicles presented in \fig{fig:lowJ} at low $\tilde{J}_I$ value (parameter values are the ones given in \fig{fig:lowJ}). Color lines are guides to the eye.}
\label{fig:SF_diffC_aboveTc}
\end{figure}

Figure~\ref{fig:SF_diffC} presents spatial composition correlation functions and respective structure factors for systems with different curvature coupling strength $c_1$ at large $\tilde{J}_I$. The correlation function shows a first peak, the width of which is proportional to pattern typical size. When the vesicle has modulated phases, it shows oscillations corresponding to pattern wavelength~\cite{Truskett}. As expected, correlation functions with larger $c_1$ have a smaller width of the first peak and a smaller wavelength, capturing small pattern size. In the structure factors, we equivalently observe that the peak position $l^*$ increases when $c_1$ increases, which leads to a smaller typical wavelength $2\pi R/l^*$ in the structured emerging patterns.

Following the definition of the structure factor $S(l)$ in \eq{Scalc}, the variance of the concentration is given by $S(0)$. We indeed measure $S(0)=0$ in our simulations, due to the fact that we have imposed a fixed concentration $\bar\phi$ (see \fig{fig:syst-size} in \app{sec:finite_size} for instance).

If the structure factor has a maximum for the first mode $l=1$, which corresponds to the soft mode $q=0$ in the planar case ($R\to\infty$) since $(Rq)^2=l(l+1)-2$~\cite{GG1}, and then decreases monotonously with $l$, the system is disordered. For low $c_1$ in \eq{eq:M}, $M(l)$ is almost quadratic in $l$, which leads to a decreasing exponential correlation function with correlation length $\sqrt{2\hat J/\hat m}R$. This is the expected Orstein-Zernicke behavior for the structure factor in the dilute phase~\cite{Chaikin}. 

The excitation of the mode $l=1$ is also maximum in the macrophase case when one hemisphere is rich in A-species and the other one in B. This corresponds to a divergence of $S(l)$ at $l=1$ in an infinite-size system. Since we consider a finite-size system, we cannot get any divergence but a maximum of large amplitude in our case.
Thus for a finite-size system $S(l)$ has a maximum at $l=1$ in both cases, disordered and macrophase states. To distinguish them, we decided to consider the ratios of amplitude between the first two modes $\rho=S(1)/S(2)$ and assumed that for $\rho \gg 1$ the structure factor corresponds to macrophase separation and for $\rho \sim 1$ to a disordered phase.

When $S(l)$ exhibits a second maximum for a value $l=l^*\neq1$, it is the signal of an underlying structuration, i.e. a modulated phase. The typical inter-domain distance is then $L=2\pi R/l^*$~\cite{Truskett}. This case corresponds to the so-called structured disordered phase (or microemulsion), as described in Sec.~\ref{sec:intro}, where composition fluctuations are stabilized by curvature coupling, as illustrated in \fig{fig:SF_diffC_aboveTc}. Theoretically, another phase has been defined \cite{GG1,Schick,Shlomo}, termed structured ordered (or mesophase, visible in \fig{fig:SF_diffC}) defined by the divergence of the structure factor for $l^*\neq1$. Again in our case, since we perform numerical simulations of a finite-size system, we cannot observe any divergence and these two phases can hardly be distinguished in practice. 

As explained in Refs.~\cite{GG1,review}, these structured disordered phases appear at large $c_1$ and small $\tilde J_I$, because the gain in bending energy is larger than the cost in line energy. The structure factor $S(l)$ exhibits a maximum for $l^*\neq1$ given by
\beq
l^*(l^*+1)-2=\tilde{q}_c^2=c_1\sqrt{\frac{\hat{\sigma}}{2\hat{J}}}-\hat{\sigma} 
\label{lstar}
\eeq 
where $\tilde{q}^2=Rq^2=l(l+1)-2$, as discussed in Ref.~\cite{GG1}. Note that $q=0$ corresponds to $l=1$ in spherical geometry.
The maximum is obtained for a non-zero value of $\tilde{q}^2$ (i.e. $l^*>1$), when $\tilde{q}_c^2>0$, which defines the threshold coupling value $c_1^*$ (see Ref.~\cite{GG1}):
\beq
c_1>c_1^*= \sqrt{2 \hat{J} \hat{\sigma}}
\label{eq:F1}
\eeq
signalling the onset of phase modulation. 

In \fig{fig:SF_diffC_aboveTc} we plotted the structure factors of the two systems presented in \fig{fig:lowJ}. Although very difficult to distinguish with the unaided eye, as expected the curvature coupling stabilizes local composition fluctuations, even above the transition, and generates underlying structuring in the vesicle mixture spatial repartition. This effect can be captured by the structure factor. Indeed, it exhibits a maximum for $l^*=1$ for the system with $c_1=0$, attesting of no structure in the corresponding system. By contrast, the structure factor of the system with $c_1=3.0$ has a maximum for $l^*=5$, related to pattern wavelength, thus revealing underlying structure. Consistently here $c_1=3.0>c_1^*\simeq1.1$, inducing phase modulation.

\begin{figure}[t]
\begin{center}
\includegraphics[width=0.85\columnwidth]{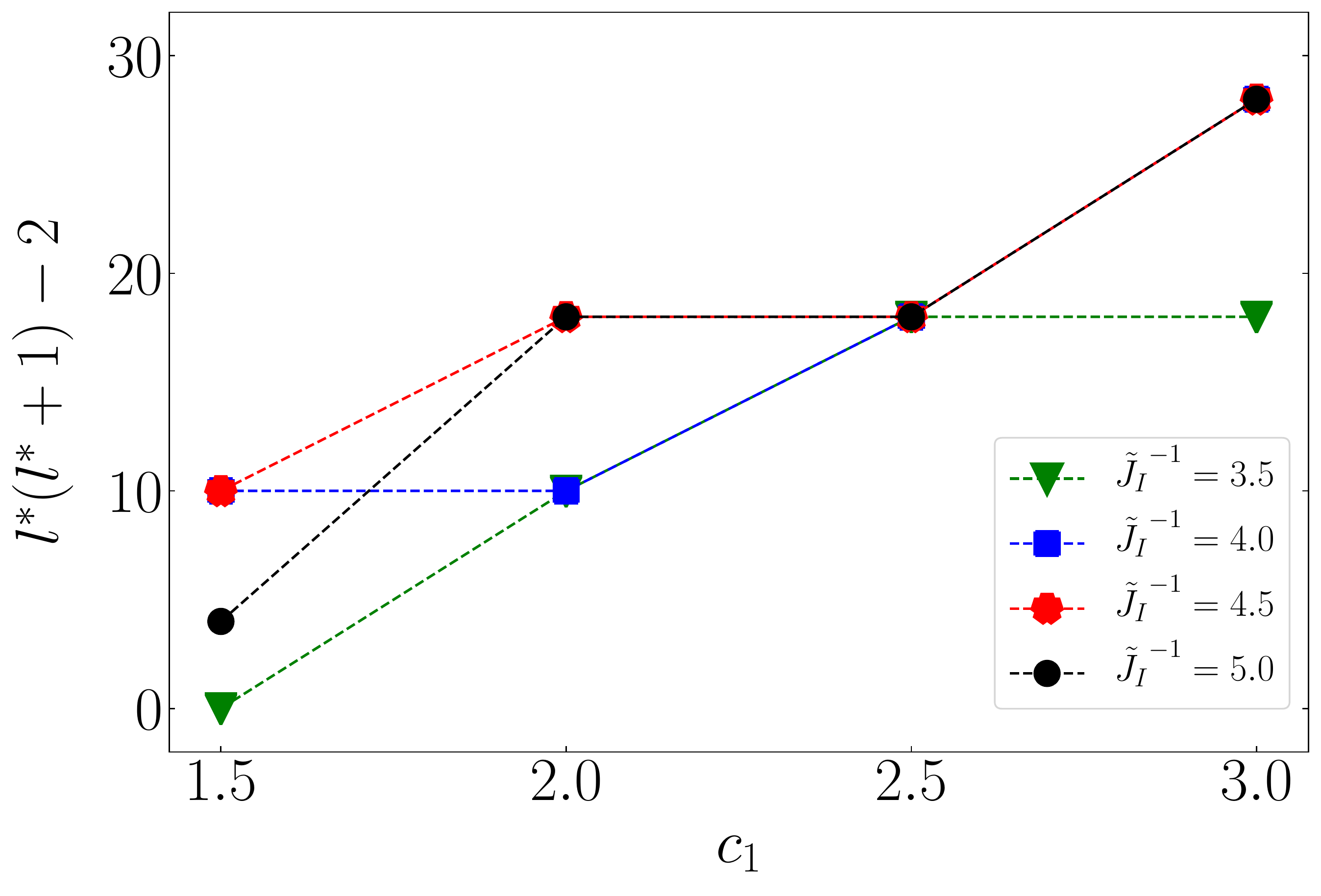} 
\caption{Variation of $l^*(l^*+1)-2$ with $c_1$, where $l^*$ is the mode corresponding to a maximum in the numerical structure factor, associated with the occurence of the structured disordered phase as shown in \fig{fig:lowJ} right.}
\label{fig:letoile}
\end{center}
\end{figure}

The $l^*$ are extracted from the numerical structure factors and shown in \fig{fig:letoile} as a function of $c_1$. They qualitatively follow \eq{lstar}, i.e. $l^*(l^*+1)-2$ grossly grows linearly with $c_1$, although it is difficult to extract any slope due to the integer values taken by $l^*$. 

We fit the numerical structure factor with the expression of $S(l)$ given in \eq{eq:SF}. The simulation parameters involved in the fit are $\hat{J}$, $\hat{m}$ and $\hat{\sigma}$.
The expression of $\hat{m}$ is related to $\hat{J}$ as given in table \ref{tab} but we miss the value of the coefficient $\alpha_0$. As explained above, a Flory mean-field approximation for this value is $\alpha_0=1/\pi$. Note that the surface tension involved in this expression is different from the input value as it is renormalized by curvature coupling and system size as explained in \app{sec:sig_renorm} (see also \refc{GG2}).

\begin{figure}[h!]
\centering
\includegraphics[width=0.97\columnwidth]{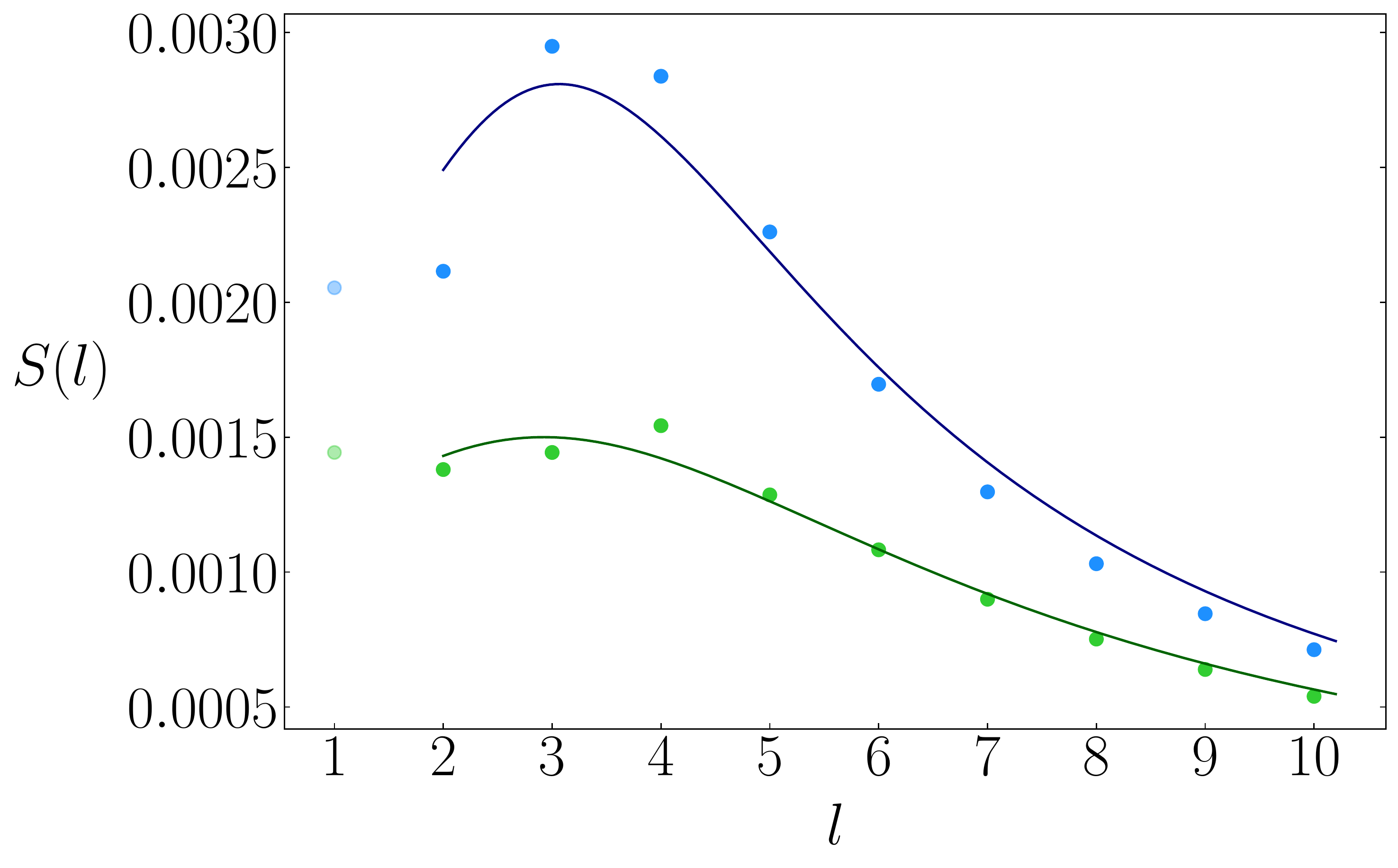} 
\caption{Fit of the structure factor for two system sizes $N=2562$ and $\tilde{\sigma}=300$ (blue), $N=10242$ and $\tilde{\sigma}=1217$ (green). Here $\tilde{J}_I^{-1}=4.0$, $\bar{\phi}=0.5$, $c_1=2.0$. The effective surface tension $\tilde{\sigma}_{\rm eff}\approx4.2$ in both cases (see \app{sec:sig_renorm}). The mode $l=1$ was not taken into account in the fitting process (see text).}
\label{fig:fits}
\end{figure}

The structure factors $S(l)$ for two different system sizes are shown in \fig{fig:fits} for a system featuring modulated phases. We observe that the structure factor amplitude is larger for $N=2562$ than for $N=10242$. This is consistent with the fact that $S(l)$ depends on $N$ via $\hat{m}=\frac{\alpha_0 N}{\tilde{\kappa}_0}(1-\frac{J_I}{J_{I,c}})$.
The expected theoretical values for $\hat{J}$ and $\hat{m}$ can be drawn from Tab.~\ref{tab}.
In \fig{fig:fits} the fitted value for $N=2562$, $\hat{J}\approx0.045$, is close to the expected one $\hat{J}_{\rm th}\approx0.043$. In contrast, the fitted value $\hat{m}\approx0.19$ differs from the excepted value $\hat{m}_{\rm th}\approx4.1$. The fitted value for $\hat{\sigma}\approx6.4$ also differs from the expected  effective surface tension $\hat{\sigma}_{\rm eff, th}\approx0.21$ (see \app{sec:sig_renorm}). This is also noticed for $N=10242$. 

The main issue is that fitting the parameters $\hat{m}$ and $\hat{\sigma}$ is very sensitive to numerical data as described in \app{sec:valley}. In the fitting process, one minimizes the squares of the distances between the theoretical values and the numerical ones. We used the GOSA software~\cite{Czaplicki, Goffe} that applies simulated annealing to fit the data. We found that the minimum is quasi-degenerate for $\hat{m}$ and $\hat{\sigma}$, in other words we have a valley of quasi-degenerate minima. This implies that if the numerical data are slightly different from the real ones, we will find strong deviations in the fitted parameter values. Another manifestation of this phenomenon are error bars on $\hat{m}$ and $\hat{\sigma}$ on the same order of magnitude as the fitted values, contrary to $\hat{J}$. Indeed, the GOSA code also provides error bars on the fitted parameters, measured during simulated annealing. Even if we were able to acquire precise fitted values of $\hat m$ and $\hat \sigma$, comparison to theory would be uneasy because of the approximations made in the mean-field calculation of the constant $\alpha_0$ as underlined above. 

Furthermore, the theory developed in \refc{GG1} is valid for infinite size systems. However, in our case, we study systems with a finite number of sites $N=2562$ or 10242. This has some consequences on system observables and especially on the structure factor. 
Above all, the first modes are affected by these finite-size effects since they correspond to large scale phenomena in real space. Note that for this reason, we did not take into account the value for $l=1$ in the fits of the structure factors.
\app{sec:finite_size} provides more detailed explanations about these finite-size effects. Increasing the system size $N$ allows us to reduce this bias and to get more accurate values for the structure factor as shown in \fig{fig:syst-size} of \app{sec:finite_size}. We then fitted the structure factor for a system of size $N=10242$ as depicted in \fig{fig:fits}.
However, we encounter numerical limitations since this system size requires considerable simulation time to have good enough sampling for the correlation measurements as mentioned in Section~\ref{sec:num_meth}. We then have reduced finite-size effects on the structure factor coefficient measurements with increased system size but poorer precision on the measured values of $S(l\geq2)$.

\subsubsection{Phase diagram at $\bar{\phi}=0.5$}
\label{sec:phase-diag}

\begin{figure}[h!]
\begin{center}
\includegraphics[width=0.96\columnwidth]{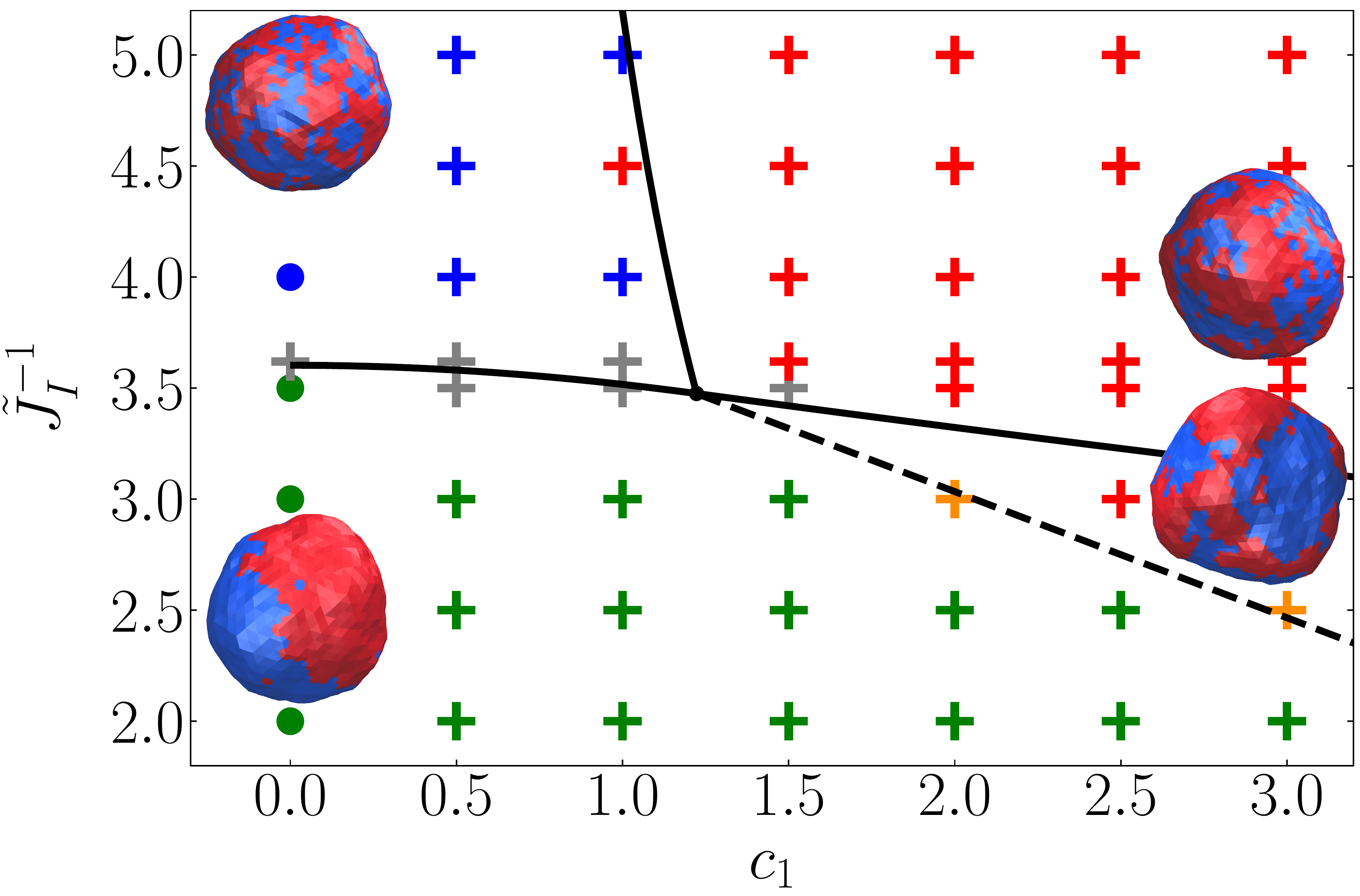} 
\caption{Phase diagram in the ($c_1$, $\tilde{J}_I^{-1}$) parameter space with $\tilde{\sigma}=300$ and $\bar{\phi}=0.5$. Green crosses are macrophases, red ones modulated phases and blue ones disordered phases. The dots for $c_1=0$ are the case without coupling (exact results~\cite{Baxter}). The solid black lines are the analytical expressions, Eqs.~\eqref{eq:F1} to \eqref{eq:F3}, of the frontiers. The dashed line separates the macrophase and the structured ordered one~\cite{GG1}.}
\label{fig:phase-diag}
\end{center}
\end{figure}

In order to study the influence of the different parameters and to compare our results to the analytical ones~\cite{GG1}, we construct a phase diagram. We focus on a 2D phase diagram in the ($c_1$, $\tilde{J}_I^{-1}$) space at fixed surface tension $\tilde{\sigma}$ and for a given composition $\bar{\phi}$. Note that these vesicles have the same input, ``bare'' surface tension in the simulations but not exactly the same effective surface tension, which is modified by the curvature coupling (see \app{sec:sig_renorm}).
The diagram shown in \fig{fig:phase-diag} compares the competing influences of the curvature coupling $c_1$ and the lipid-lipid affinity through the Ising parameter $\tilde{J}_I$. We clearly distinguish three regions instead of two in a classical Ising system without coupling:
\begin{itemize}
\item a disordered region for low $\tilde{J}_I$ and $c_1$ values, where the mixture is homogeneous and features no underlying order (blue crosses);
\item a macrophase region for high $\tilde{J}_I$ values and low $c_1$ coupling, in which the lipid mixture undergoes complete macrophase separation (green crosses);
\item a modulated phase region for large $c_1$ and low $\tilde{J}_I$, in which the vesicles feature more than one domain and where the mixture is then modulated (red crosses). This region contains the numerically indistinguishable structured disordered (low $\tilde{J}_I$) and structured ordered (high $\tilde{J}_I$) regions described above.

\end{itemize}

The vesicle states are classified into these different phases via the observation of the position of the structure factor maximum and its amplitude (see section \ref{sec:SF}). In some cases, the distinction is unclear because the maximum position is difficult to determine accurately enough due to measurement errors and discretisation. Moreover the phase determination has to take into account the finite size of our system, since $S(l)$ has a maximum in both cases disordered and macrophase. To distinguish them, we thus used the criteria for $\rho$ set in \ref{sec:SF}. When the value of $\rho$ was in-between, we depicted this with grey crosses. For some systems in the vicinity of the frontier between macrophases (dominant peak at $l=1$) and structure ordered phases (dominant peak at $l=l^*>1$), the peaks at $l=1$ and at $l=l^*$ have comparable heights. We assigned orange crosses to these particular borderline cases.

We derive the analytical expressions of the region frontiers in the phase diagram. The expressions of $M(l)$ and $S(l)$ are given respectively by Eqs.~(\ref{eq:M}) and (\ref{eq:SF}). The equation of the frontier in the phase diagram between disordered (blue crosses) and structured disordered phases (red crosses) is $c_1=c_1^*$ where $c_1^*$ is given in \eq{eq:F1}.

As already mentioned in \ref{sec:SF}, the structure factor diverges at $\tilde{q}_c^2=0$ when $M(1)=0$ leading to the equation
\beq
\hat{m}+c_1^2+4\hat{J}=0
\label{eq:F2}
\eeq
giving the frontier equation between macrophase separation and disordered phase (green/blue, solid line on the left side of the triple point).

The equation of the frontier between structured disordered and structured ordered phases (solid line on the right side of the triple point) is obtained  when $S$ diverges for $\tilde{q}_c^2\neq0$ which leads to 
\beq
\hat{m}+2c^*_1c_1-c_1^{*2}+4\hat{J}=0.
\label{eq:F3}
\eeq
Again this theoretical frontier cannot be identified numerically for a finite-size system. In \app{sec:front} are derived the exact expressions of these frontiers in terms of $\tilde J_I$ and $c_1$.

For higher $\tilde{J}_I$, the Gaussian Hamiltonian is no more valid and a term in $\phi^4$ should be kept in the theory. This is beyond the scope of this work, thus the frontier between macrophase and structured ordered phases is determined only numerically and shown with a dashed line in \fig{fig:phase-diag}. 

\subsubsection{Phase diagram at $\bar{\phi}=0.2$}

We also study vesicles at $\bar{\phi}=0.2$\footnote{We have also measured numerically at $\bar{\phi}=0.2$ that the specific heat of the pure Ising system exhibits a maximum at $\tilde{J}_{I,c}^{-1}\simeq3.45$. This is consistent with the fact that at $\bar{\phi}\neq \bar{\phi}_c$ the system undergoes a first-order transition at a higher $\tilde{J}_I$ than at $\bar{\phi}=\bar{\phi}_c$~\cite{Chaikin}.}, a concentration that is more illustrative of biological membranes containing curvature-generating proteins or particular lipids~\cite{Phillips}.

\begin{figure}[t]
\centering
\includegraphics[width=0.5\columnwidth]{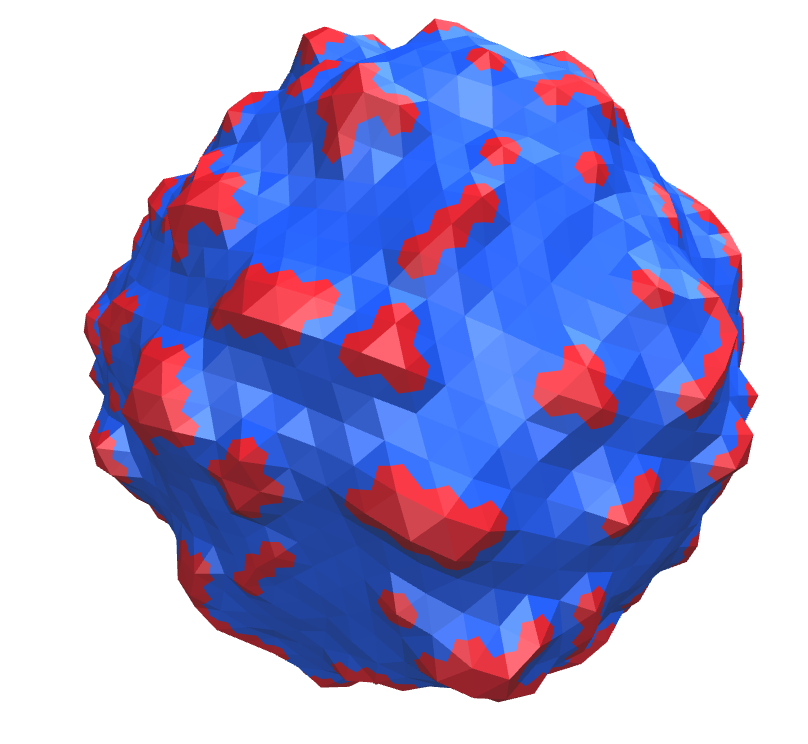} 
\caption{Snapshot of a simulated vesicle for $\bar{\phi}=0.2$ and $c_1=15$. ($\tilde{J}_I^{-1}=2.5$ and $\tilde{\sigma}=300$). Numerous small curved domains are observed, as expected.}
\label{fig:c17}
\end{figure}

\begin{figure}[t]
\centering
\includegraphics[width=0.96\columnwidth]{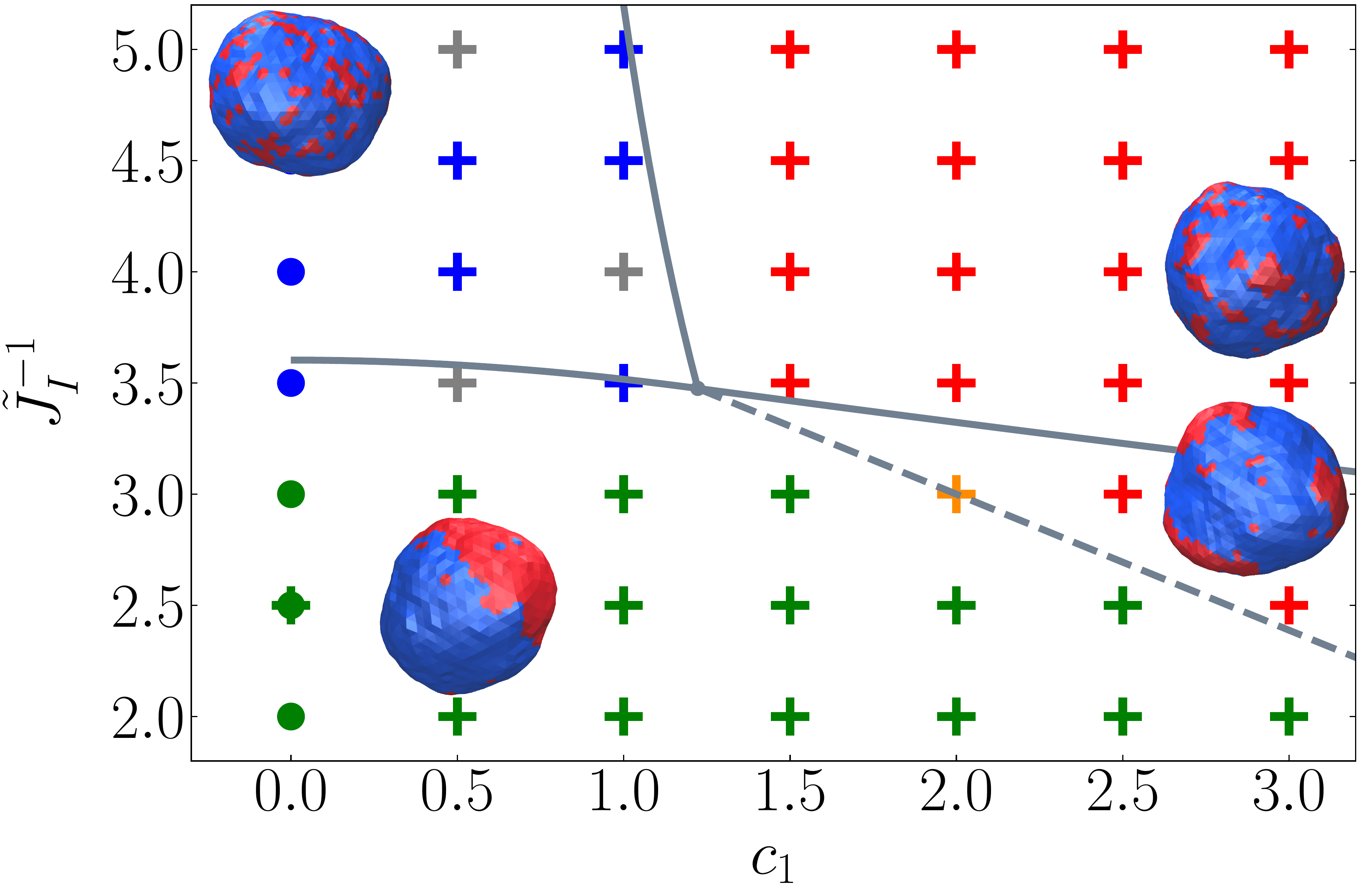} 
\caption{Phase diagram in the ($c_1$, $\tilde{J}_I^{-1}$) parameter space with $\tilde{\sigma}=300$ and $\bar{\phi}=0.2$. The color code is the same as in \fig{fig:phase-diag}. The grey lines are Eqs.~(\ref{eq:F1}-\ref{eq:F3}) plotted with the parameter values for $\bar{\phi}=0.5$, see Fig.~\ref{fig:phase-diag}.}
\label{fig:phase-diag02}
\end{figure}

In \fig{fig:c17} is shown a simulated vesicle for $\bar{\phi}=0.2$ and high coupling value, $c_1=15$. As explained in Section \ref{sec:effetC}, we observe many small curved domains. We also built the same phase diagram as in \fig{fig:phase-diag} but at $\bar{\phi}=0.2$, as shown in \fig{fig:phase-diag02}. There is no reason why the phase diagrams at $\bar{\phi}=0.2$ and $\bar{\phi}=0.5$ should coincide. However, we observe that they are very similar, except in the close vicinity of the frontiers. The grey lines in \fig{fig:phase-diag02} are the same as the ones in \fig{fig:phase-diag}, derived from the theory at $\phi_c=1/2$, and are just a guide to the eye. 

\subsubsection{Domain number and size distribution}

Above $\tilde{J}_{I,c}$ we can also characterize the emerging domains in terms of number and size. Cluster detection analysis is performed in order to compare cluster size distributions only for low enough $\bar{\phi}$ to have well-defined clusters. We recall that at high concentration domains merge into labyrinthine structures percolating through the vesicle, as visible in \fig{fig:effetC05}, right-most vesicles. Hence at $\bar\phi=0.5$, mesophases and macrophases are hardly distinguishable close to the dashed line using the sole cluster size distribution.

\begin{figure}[t]
\begin{center}
\includegraphics[width=0.93\columnwidth]{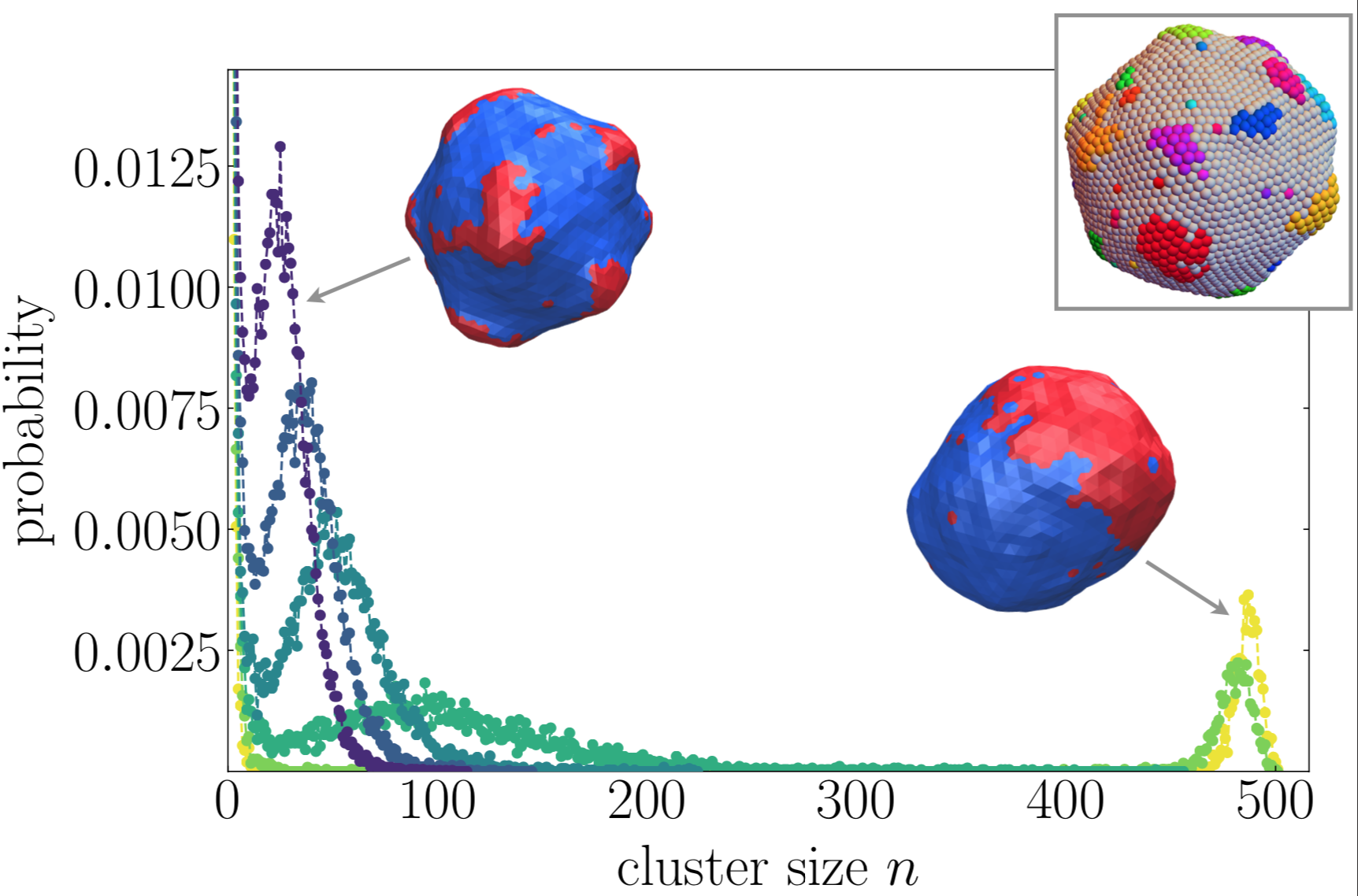} 
\caption{Cluster size distributions $p(n)$ for  $c_1$ increasing from 1 to 6 (from yellow to purple). We can distinguish macrophases (right) and mesophases (left) via the peak abscissa which corresponds to the typical cluster size $n^*$. The size of the clusters is measured in units of number of elements. $\bar{\phi}=0.2$, $\tilde{J}_I^{-1}=2.5$ and $\tilde{\sigma}=300$. Color lines are guides to the eye.) Inset: Cluster detection analysis performed on a simulated vesicle with the DFS algorithm. Each color corresponds to an identified cluster.}
\label{fig:size_dist}
\end{center}
\end{figure}

At $\bar\phi=0.2$, we implemented a depth-first search (DFS) algorithm in order to identify the different clusters and to index their size in units of number of sites. We then plot the size distribution, i.e. the occurrence $p(n)$ of clusters comprising $n$ sites throughout the simulation, after equilibration, as shown in \fig{fig:size_dist} for various values of $c_1$. These distributions show a local maximum at the most probable cluster size. In the case of cluster phases, the distribution is bimodal and the secondary peak position $n^*$ corresponds to the typical domain size, measured in units of the number of sites belonging to a same cluster. To get accurate values of $n^*$ we fit the secondary peak with a Gaussian. Note that in the case of a macrophase, the distribution shows a peak the abscissa of which is close to the total number of A-species, as observed in \fig{fig:size_dist}, coherent with the fact that most of the A sites are condensed in a single macro-cluster. 

On a triangular lattice, the typical cluster size $n^*$, the typical inter-cluster distance $L$ and the position $l^*$ of the structure factor maximum are related through
\beq
n^*=\frac{L^2}{a^2}\bar{\phi}  = \left(\frac{2\pi R}{a l^*}\right)^2\bar{\phi}  = \frac{\sqrt{3} \pi}2 \frac{N}{{l^*}^2}\bar{\phi}
\label{eq:size}
\eeq
owing to $L=2\pi R/l^*$, $a$ still being the lattice parameter.

For example for the case shown in \fig{fig:c17}, we find that $l^*=16$. Using \eq{eq:size} leads to a typical cluster size $n^*$ of 5 sites, which is also the size found using the cluster size distribution secondary peak position. Both approaches are mutually consistent. In a real vesicle with a radius of $10~\mu$m, these domains would have a diameter on the order of $1~\mu$m, the same order of magnitude as the curvature-induced lipid domains observed experimentally in~\refc{Shimobayashi}.

We also studied the effect of $\tilde{J}_I$ on domain formation using the size distributions (data not shown). At high enough $c_1$ coupling, the system features domains getting greater as $\tilde{J}_I$ increases and even fuse into a macrophase when $\tilde{J}_I$ is high enough, corresponding to the region below the dashed line in \fig{fig:phase-diag02}. 
The clusters coexist with a population of low-density, dispersed monomers and small multimers, the so-called gas phase. The clusters continuously exchange monomers with this homogeneous gas phase. This can be seen as an analogue of a liquid-gas coexistence~\cite{Truskett}.
While $\tilde{J}_I$ increases, the monomers become increasingly scarce and nearly all condensed into clusters because the line tension is very high and their detachment has a high cost in terms of interfacial energy.
The first peak of the distribution $p(n)$, corresponding to monomers and small multimers, seems to be well fitted by a power-law, which might be explained by the reminiscence of the critical behavior of the Ising model in the vicinity of the critical point~\cite{Toral}.

\begin{figure}[t]
\centering
\includegraphics[width=0.83\columnwidth]{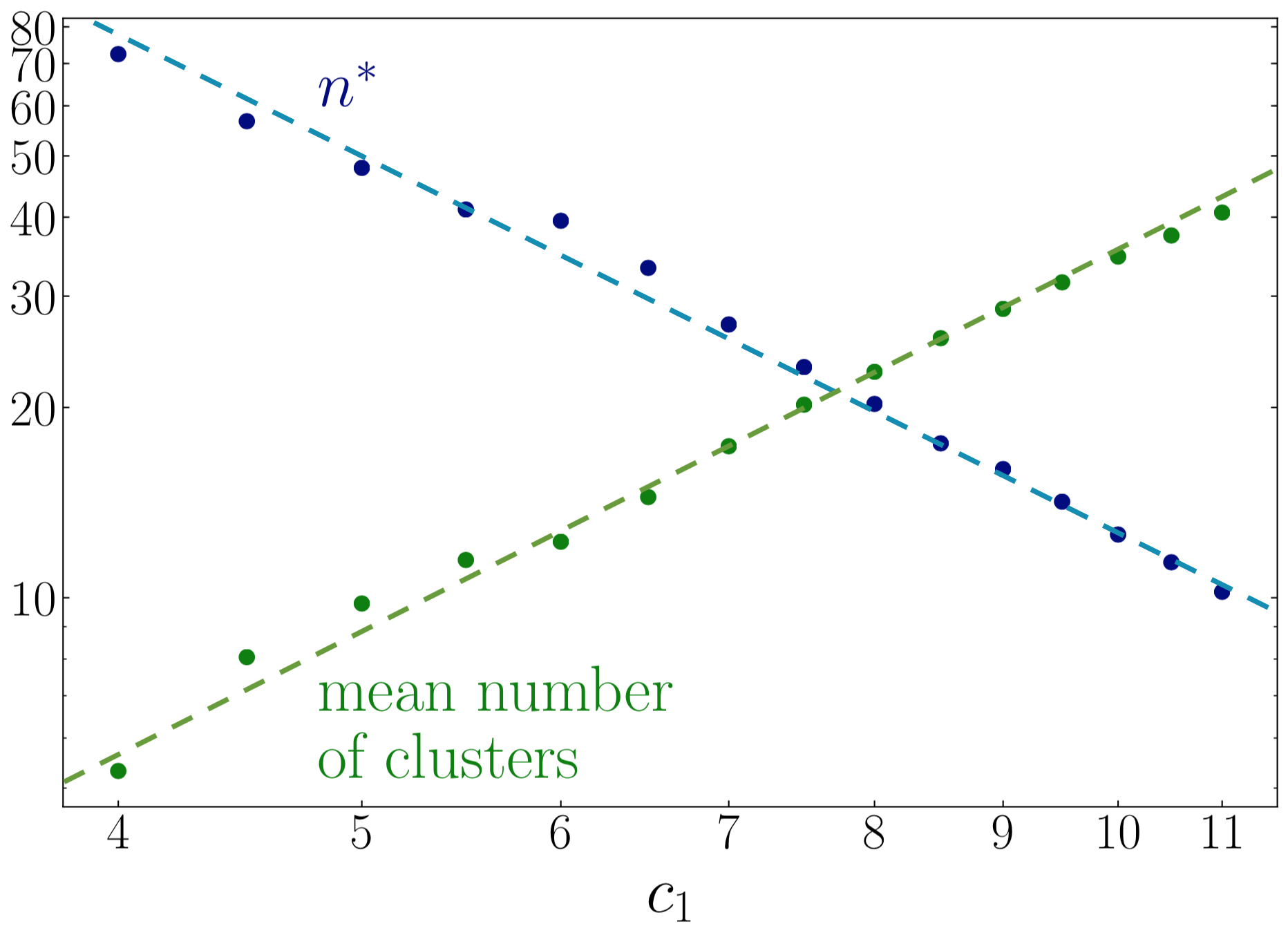} 
\caption{Effect of the curvature coupling $c_1$ on the typical cluster size $n^*$ extracted from \fig{fig:size_dist} and on the mean number of clusters for clusters of size $n\geq5$ ($\bar{\phi}=0.2$, $\tilde{J}_I^{-1}=2.0$ and $\tilde{\sigma}=300$). Log-log coordinates. The dashes lines have slope 2 and -2, respectively, and are guides to the eye.}
\label{fig:clust_size_num}
\end{figure}

We now study the domain typical size and their number as a function of the curvature coupling $c_1$ in \fig{fig:clust_size_num}. 
As described above, we see that the increase of the coupling leads to the formation of smaller (blue points) and then more numerous curved membrane domains (green points).

In \fig{fig:clust_size_num} we note that the typical cluster size (area) $n^*$ scales with $c_1$ with a power law of $-2$ exponent. The authors of~\refc{Shimobayashi} recently found an experimental $-1$ exponent for domains induced by curvature-generating externally added glycolipids in GUVs. However their data have significant error bars and this exponent will have to be confirmed in future experiments. We also observe that the average number of clusters roughly scales like $c_1^2$. Together with the scaling $n^* \propto c_1^{-2}$ discussed above, we find that the total number of A sites in clusters is almost constant, as expected at high $\tilde{J}_I$ where most of A sites are condensed into clusters.

\section{Conclusion and discussion}

The curvature-composition coupling mechanism studied in this work is a good candidate to explain the formation of certain mesodomains in biomembranes.
In \refc{review} we argued that it is one of the few most suitable mechanisms to explain the existence of membrane domains whose size is shorter than optical resolution, provided that some molecular species break the up/down symmetry of the membrane by curving it sufficiently. Below the demixing temperature (strong segregation limit), aggregation is stopped before a macrophase emerges because curvature makes too large domains unstable. Above the demixing temperature (weak segregation limit), density fluctuations have a typical size
corresponding to a maximum of the structure factor, whereas their size distribution would decay exponentially without any coupling to curvature (Orstein-Zernicke behavior in a dilute phase~\cite{Chaikin}). In both cases, density and shape fluctuations are coupled~\cite{review}.

We were able to validate numerically the theoretical phase diagram proposed in Ref.~\cite{GG1}, in particular  above $J_{I,c}$ (i.e. in the strong segregation limit) where the approximate theoretical developments were not expected to hold. In addition to the modulation wavelength that is accessible through the maximum of the structure factor, we calculated the typical cluster size in the low-concentration region were clusters are well-defined. We checked that both approaches are mutually consistent. Contrary to the analytical single mode approximation used in Refs.~\cite{Hansen1998, Kumar1999, Jiang2000, Harden2005}, one observes qualitatively, by looking at the snapshots, and also quantatively through the structure factor, that the domains are actually far from being stripes or periodically organized roundish domains. One rather obtains deformed domains with a liquid (e.g. short range) order.

Concerning the question raised in the Introduction of the experimentally observed domain sizes, below the diffraction limit, we proposed a scaling law for the typical cluster size $n^*$ in function of the spontaneous curvature $c_1$ of the minority species, $n^* \simeq 1200/c_1^{2}$ (for $N=2562$, as shown in \fig{fig:clust_size_num}). This means that cluster radii are $r=2R\sqrt{n^*/N}\simeq 1.4/C_1$ in the studied regime of parameters. We could not go beyond a limiting value of $C_1$ because $r$ would have become comparable or even smaller than the lattice parameter $a\simeq \sqrt{4\pi R^2/N}\simeq 700$~nm for a vesicle radius $R=10~\mu$m. However, we can extrapolate the scaling law $r \simeq 1.4/C_1$ beyond the simulated values. A typical size $r \approx 50$~nm, commonly observed by super-resolution microscopy, would lead to $C_1 \approx 0.03$~nm$^{-1}$. This value is readily attainable for lipid~\cite{Hossein2020,Perlmutter2011,review} or protein domains~\cite{Zimmerberg2006}. As a consequence, experimentally observed domain sizes can be accounted for by the model presented in this work~\cite{review}.

A limit of numerical methods is that they can only address finite-size systems and cannot tackle rigorously phase transitions characterized by singularities of physical observables in the thermodynamic limit. In particular, two kinds of modulated phases can appear in our system, as discussed thoroughly in Ref.~\cite{GG1}: the so-called ``structured disordered'' phase (or microemulsion) above the demixing temperature and the ``structured ordered'' phase (or mesophase) below the demixing temperature. These two phases can only be distinguished through their structure factors. In the first case, it has a maximum at finite wave-vector, which becomes a divergence in the second case. In a finite-size systems, both phases simply present a maximum and become indiscernible. We cannot easily conclude on the precise frontier between both phases. 

The next step of this work will be to address models where not only the spontaneous curvature depends on local concentration, but also the bending rigidity $\kappa$, because the membrane thickness depends on its phase state. Some numerical works have tackled this issue (see Ref.~\cite{review} for a review), but no systematic study has explored the corresponding phase diagrams and the entanglement between spontaneous curvature and bending modulus. From a numerical perspective, this leads to consider a 4-state Potts model coupled to the membrane shape to explicitly deal with the two membrane leaflets, which leads to much more complex phase diagrams~\cite{GG1}. 

We have not studied in detail the transition from roundish domains to labyrinthine phases either, as observed between $\bar \phi=0.2$ and 0.5. Tension has also been demonstrated to play a role in this context~\cite{Komura2006}. This morphological transition can be of particular biophysical interest because membrane cell domains are often supposed to be disjoint. However, in some experiments, some proteins are highly over-expressed to get a strong enough fluorescent signal. This might lead to an undesired change in domain morphologies, a possible experimental bias that should be quantified in the future.

\begin{acknowledgments}
We warmly thank Guillaume Gueguen for his valuable work on the simulation code and his precious help. We are indebted to Matthieu Chavent, Adrien Schahl and Sarah Veatch for fruitful discussions. We also warmly thank Georges Czaplicki for kindly helping us in handling the GOSA software. 
\end{acknowledgments}

\appendix

\section{Summary of previously published results}

\subsection{Structure factor}
\label{SF}

In the previous analytical work~\cite{GG1}, the total quadratic Hamiltonian $H[u,\phi]$ is written as the sum of 3 contributions:
\begin{itemize}
  \item $H_{\rm Helf}[u]$, the Helfrich Hamiltonian describing height fluctuations and membrane elasticity;
  \item $H_{\rm GL}[\phi]$, the Ginzburg-Landau Hamiltonian accounting for lipid-lipid (or protein-lipid) interactions in the binary mixture;
  \item $\delta H[u,\phi]$, the coupling contribution.
\end{itemize}
In order to study the structure factor, $H$ is written in the spherical harmonics basis. The height function $u$ writes
\beq
u(\theta, \varphi)=\frac{u_{00}}{\sqrt{4\pi}}+ \sum_\lambda u_{lm} Y^m_l(\theta, \varphi)
\eeq
where $\sum_\lambda=\sum^{l_{\rm max}}_{l=1}\sum^l_{m=-l}$ with $l_{\rm max}$ the ultraviolet cutoff. The same holds for $\phi$, with coefficients $\phi_{lm}$.
We recall that the spherical  harmonics are defined as~\cite{Abramowitz}
\beq
Y^m_l(\theta, \varphi)= (-1)^m \sqrt{\frac{2l+1}{4 \pi}\frac{(l-m)!}{(l+m)!}} P_l^m(\cos \theta) e^{im\varphi}
\eeq
where $P_l^m$ are the Legendre functions here defined as
\beq
P_l^m(x)= \sqrt{(1-x^2)^m} \frac{{\rm d}^m}{{\rm d}x^m}P_l(x) \qquad (-1 \leq x \leq 1)
\eeq
with $P_l$ the Legendre polynomials. $H_{\rm Helf}[u]$, $H_{\rm GL}[\phi]$ are written in this new basis, where they are now diagonal quadratic forms, 
of respective diagonal coefficients $H_{\rm Helf}(l)= \frac{\kappa_0}2 [l(l+1)-2] \left[ l(l+1) + \hat{\sigma} - c_0 \left( 2 - \frac{c_0}2 \right) \right] $ and $H_{\rm GL}(l)=\frac{m}2+Jl(l+1)$. The term $\delta H[u,\phi]$ becomes $\delta H (l)=\Lambda[l(l+1)+2-2c_0]$ that couples the $u_{lm}$ and $\phi_{lm}$. We recall that $\Lambda=-\kappa_0 C_1$ is the coupling constant between the concentration field and the local curvature.

The quadratic Hamiltonian $H[u,\phi]$ can now be integrated on $u$ which yields~\cite{GG1}  the structure factor of the composition field $\phi$
\beq
S(l)=\frac{k_{\rm B}T}{2 \pi \kappa_0} \frac{1}{M(l)}
\eeq
where $M(l)$ is given in Eq.~\eqref{eq:M} in the case $c_0=2$.

Note that the vesicle description being isotropic, it is independent of the spherical coordinate $\varphi$ and $m=0$ in the $Y^m_l$ description.

\subsection{Surface tension renormalization by curvature coupling and system size}
\label{sec:sig_renorm}

Following \refc{GG2}, for uniform spontaneous curvature $c$, the effective surface tension depends on $c$ and on the number of sites $N$ as
\beq
\tilde{\sigma}_\text{eff}=\tilde{\sigma}+\frac{1}2\tilde{\kappa}_0(2-c)^2-\epsilon \frac{N}{8\pi}
\label{eq:sig_renorm}
\eeq
$\tilde{\sigma}$ being the input (``bare'') surface tension in the simulation. Eq.~\eqref{eq:sig_renorm} has been found by doing renormalization calculations of  $u^4$ terms in \refc{GG2}. In this work the local curvature $c$ did not depend on $\phi$. We then write here a mean-field extension of this expression by using the mean value of the curvature $\bar{c}=c_0+c_1 \bar{\phi}$. 

Furthermore the value of the numerical prefactor $\epsilon$ depends on the type of vertex elementary moves chosen in the Metropolis algorithm. For radial moves used in our work we have $\epsilon=3$.
If in addition $c_0=2$, then we get 
\beq
\tilde{\sigma}_\text{eff}=\tilde{\sigma}+\frac{1}2 \tilde{\kappa}_0 c_1^2\phi^2 -\frac{3N}{8\pi}
\eeq
Note that this is a rough estimate for inhomogeneous membrane composition. A more precise expression should be obtained using renormalization calculations in future works.

\section{Reduced parameters in Ising and Landau models}
\label{sec:link}

We now make the connection between the parameters of the discrete Ising (lattice gas) model on a triangular lattice, and those of the continuous Ginzburg-Landau theory. 

The interaction energy between nearest-neighbor sites of the hexagonal lattice in the Ising model is
\begin{equation}
H(\{s_i\}) = -J_{\rm I} \sum_{\langle i,j \rangle} s_i s_j
\end{equation}
with $s_i = \pm 1$ and $J_{\rm I}>0$. We remind that the sum runs on $N$ lattice vertices, most of which have $\nu = 6$ nearest neighbors.

At the Gaussian order, valid below the critical Ising parameter $\tilde{J}_{I,c}$, the continuous field theory is 
\begin{equation}
H[\phi] = \int {\rm d}S \left[ \frac{m}2 (\phi-\phi_c)^2 + \frac{b}2 (\nabla \phi)^2 \right]
\end{equation}
where $\phi(\mathbf{r}) \in [0,1]$ is the composition field, $\phi_c =1/2$ is the critical composition, $m>0$ is the theory ``mass'' and $b>0$ is its stiffness.

In \refc{GG1} $b$ is denoted by $2J$ (not to be confused with $J_{\rm I}$ above), the factor 2 coming from the fact that there are initially two composition fields, one for each leaflet. The Ginzburg-Landau parameter playing the same role as our $m$ is $m_-$ but we shall denote it as $m$ because they have the same meaning. In \refc{GG1}, dimensionless quantities are introduced, namely 
\begin{equation}
\hat J \equiv \frac{J}{\kappa_0} = \frac{b}{2\kappa_0}~~\mbox{ and }~~\hat m \equiv \frac{m R^2}{\kappa_0}
\end{equation}
We want to express these quantities in function of our model parameters. 

We first relate $\phi$ and $s_i$ through $\phi_i=\phi( \mathbf{r}_i) =(1+s_i)/2 $.
In the tessellation, we consider an elementary triangle of vertices denoted by $i$, $j$ and $k$ bearing the three compositions $\phi_i$, $\phi_j$ and $\phi_k \in \{0,1\}$. We identify $\|\nabla \phi\|$ with the slope of the plane defined by the points $(i,\phi_i)$, $(j,\phi_j)$ and $(k,\phi_k)$. After a short calculation, one gets
\beq
\|\nabla \phi \|^2 =  \frac{4}{3 a^2} \left(\phi_i^2+\phi_j^2+\phi_k^2\right)^2 - \frac{4}{3 a^2} (\phi_i \phi_j + \phi_i \phi_k + \phi_j \phi_k)
\eeq
where $a \propto R/\sqrt{N}$ is again the lattice spacing. The elementary triangle has average area $\sqrt{3} a^2 /4 $. Thus skipping irrelevant squares $\phi_i^2$ of trivial integral, we obtain
\bea
\frac{b}2 \int(\nabla \phi)^2  {\rm d}S &\simeq& -\frac{b}2 \sum_{\rm triangles}  \frac{\sqrt{3} a^2}{4} \frac{4}{3 a^2} (\phi_i \phi_j + \phi_i \phi_k + \phi_j \phi_k) \nonumber \\
&=& - \frac{b}{\sqrt{3}} \sum_{\langle i,j \rangle} \phi_i \phi_j,
\eea
where a factor 2 arises from the fact that each triangle edge belongs to two elementary triangles. Owing to the relation $\phi_i =(1+s_i)/2 $, and now skipping irrelevant linear terms, we finally conclude that $b = 4 \sqrt{3} J_{\rm I}$. It follows that 
\beq
\hat{J}=\frac{2\sqrt{3}J_I}{\kappa_0}.
\label{hatJ}
\eeq
Alternatively, we could have used a more rigorous, but more technical, Hubbard-Stratonovitch transformation to reach the same conclusion~\cite{Alastuey}.

As far as $\hat m$ is concerned, we need the expression of the ``mass'' $m$ of the Ising model on a triangular lattice, assuming that the 12 sites of coordination number 5 are negligible in the large $N$ limit. 
The Ising energy is an intensive quantity and thus scales as $N$. The contribution of $m$ in the Ginzburg-Landau energy is proportional to the surface and then scales as $mR^2$, with $R$ fixed in our simulations. Combining these scalings implies that $m$ has to scale as $N/R^2$. In addition, we know that $m \propto(1 - J_I/J_{I,c})$~\cite{Chaikin}. It follows that $m = \alpha_0 \, k_{\rm B} T \, (1 - J_I/J_{I,c}) \, N / R^2$ where $\alpha_0>0$ is a non-universal dimensionless coefficient  and thus that 
\beq
\hat m = \alpha_0 \frac{N}{ \tilde{\kappa}_0} \left(1 - \frac{J_I}{J_{I,c}}\right)
\label{hatm}
\eeq
The Flory theory leads to $\alpha_0=1/\pi$. This value is only a rough estimate because this mean-field theory is not rigorous close to the critical point.

\section{Sphere tessellation bias}
\label{sec:tess}

To create the discretized sphere for the initial configuration, we start from a regular icosahedron. We then subdivide each face of the starting icosahedron by joining the middles of its 3 vertices. We then get 4 smaller triangles into each face and reiterate this process. This leads to accessible system sizes $N=10 \times 4^p+2$ where $p$ is the number of subdivision iterations (see \refc{GG2}).
A first point to notice is that this discretisation also leads to a few defects in the structure. Indeed, the connectivity of all the vertices is not equal to 6 for all of them, because of the 12 vertices of the initial icosahedron that only have 5 neighbors. This feature is taken into account when computing the local energy (Helfrich or Ising) of a vertex, however it might induce small local errors.

The main problem of this tessellation method does not come from these 12 5-neighbor vertices, but from a global issue: after each subdivision, the newly created points are projected on the sphere. The triangles close to the centers of the initial icosahedron faces then have a larger surface than the ones close to the initial vertices. This results in triangles with different areas in the sphere, the largest triangles being typically 10\% larger than the smallest ones.
The Vorono\"i area associated to each vertex~\cite{GG2} are then also different. The bending energy of a vertex is proportional to the Vorono\"i area associated with it. Thus the most curved A-species regions tend to get anchored to the smallest triangles, close to the 12 initial vertices, which biases the free energy minimisation.
To correct this effect, we tried to create an initial discretized sphere with all triangles of equal size. Since it appears to be an open problem~\cite{Harrison}, we tackled this problem numerically. Starting from the configuration generated by the above-described tessellation, we use a Metropolis algorithm at zero temperature to minimize the standard deviation of the lattice triangle areas, the local moves of which are small displacements of the vertices on the sphere. We obtain a highly peaked distribution around a characteristic triangle area, having reduced triangle area dispersion by a factor $\sim 100$.

\begin{figure}[t]
\centering
\includegraphics[width=0.9\columnwidth]{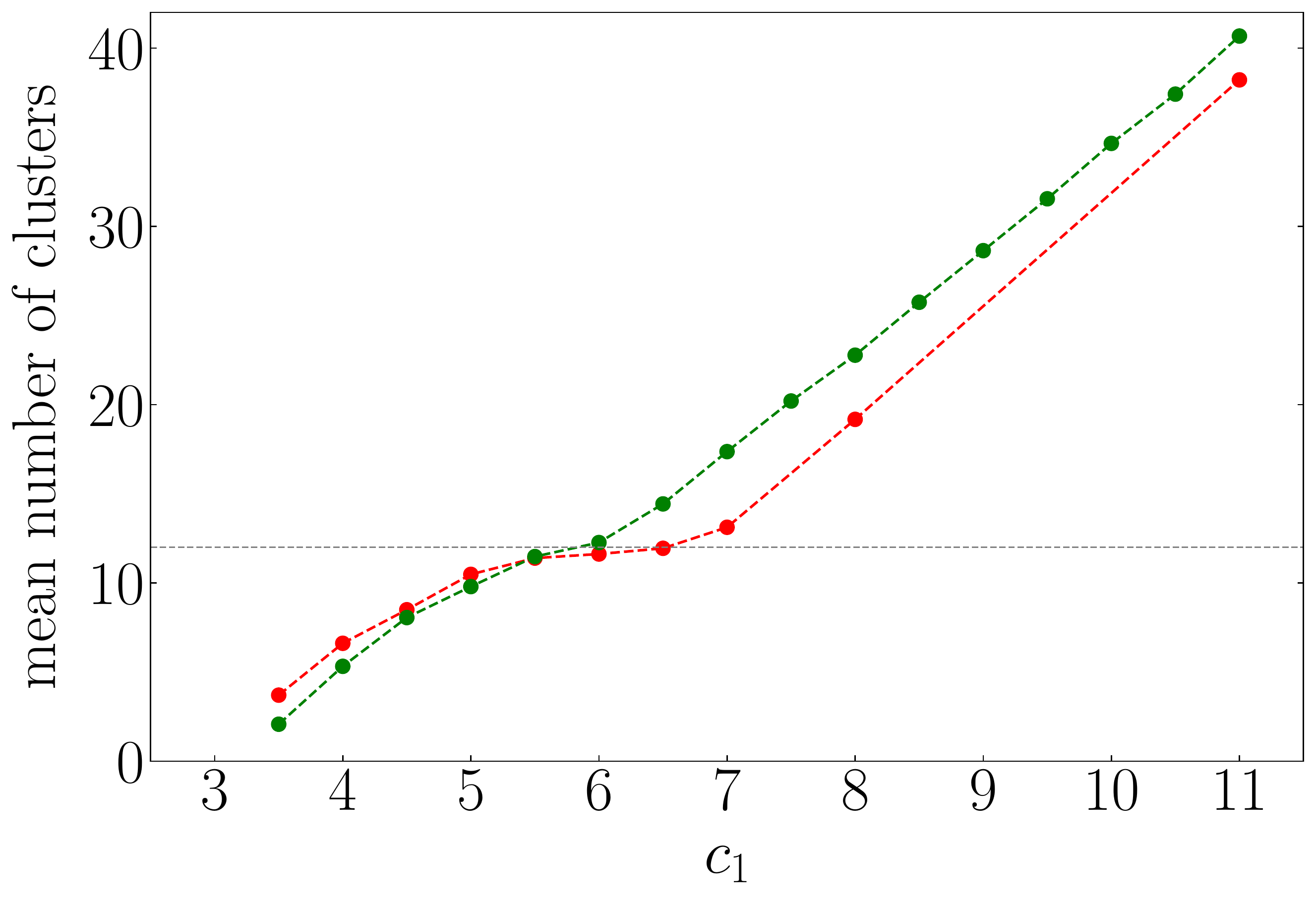} 
\caption{Mean cluster number vs curvature coupling $c_1$ for clusters of size $n\geq5$ ($\bar{\phi}=0.2$, $\tilde\sigma=300$ and $\tilde J_I^{-1}=2$) with (green) and without (red) correction of the initial triangle area distribution. Lines are guides to the eye.}
\label{fig:clustnum_cor}
\end{figure}

The bias induced by the original tessellation is illustrated in \fig{fig:clustnum_cor} by the mean number of domains with respect to the curvature coupling $c_1$.
Around $c_1=5.5$, there is a marked shoulder corresponding to systems with mean number of domains close to 12 without correction (in red). The vesicles with a little less and a little more than 12 clusters seem to be constrained to have 12 clusters because curved domains are anchored to the 12 icosahedron vertices. After correction, this shoulder almost disappeared (in green), showing that we significantly improved the triangle area distribution.

\section{Fits of the structure factors}
\label{sec:valley}

When fitting the structure factors, we minimize the square of the distance between the theoretical expression of $S(l)$ in \eq{eq:SF} and the numerical data $S_{\rm num}(l)$,  denoted by $d^2$. We used the GOSA software~\cite{Czaplicki, Goffe} that performs simulated annealing. As explained in the main text, we obtained good fitted values of the parameter $\hat{J}$, with small error bars (provided by the software). In contrast, the fitted values of $\hat{m}$ and $\hat{\sigma}$ were quite far from the expected ones, with large error bars, suggesting that $d^2$ has a degenerate minimum in the parameter set. This can be understood thanks to the  following argument.

The squared distance $d^2$ is defined as
\begin{equation}
d^2(\hat{J},\hat{m},\hat{\sigma}) = \sum_{l \geq 2} \left[ S(l) -  S_{\rm num}(l) \right]^2. 
\label{first:deriv}
\end{equation}
The data $S_{\rm num}(l)$ are fixed and we look for the the parameter values of the variables $\hat{J}$, $\hat{m}$, and $\hat{\sigma}$, appearing implicitly in $S(l)$ that minimize $d^2$.
Denoting generically these variables as $x_i$, we have the partial derivatives
\begin{equation}
\frac{\partial d^2}{\partial x_i} = 2 \sum_{l \geq 2}  \frac{\partial S(l)}{\partial x_i}   \left[ S(l) -  S_{\rm num}(l) \right]
\label{scnd:deriv}
\end{equation}
that must vanish at the minimum of $d^2$. We also need the second order derivatives to characterize this minimum: 
\begin{equation}
\frac{\partial^2 d^2}{\partial x_i \partial x_j} = 2 \sum_{l \geq 2} \left\{ \frac{\partial^2 S(l)}{\partial x_i \partial x_j}   \left[ S(l) -  S_{\rm num}(l) \right]
+ \frac{\partial S(l)}{\partial x_i} \frac{\partial S(l)}{\partial x_j} \right\}.
\end{equation}
We focus on the most favorable case where the measured values are close to the exact ones, in which case $S(l) \simeq S_{\rm num}(l) $ at the minimum of $d^2$. It follows that the first derivatives in Eq.~\eqref{first:deriv} vanish, as expected. Owing to Eq.~\eqref{eq:SF}, the second derivatives become
\begin{equation}
\frac{\partial^2 d^2}{\partial x_i \partial x_j} = \frac2{\pi^2 \tilde \kappa_0^2} \sum_{l \geq 2}  \frac1{M^4(l)} \frac{\partial M(l)}{\partial x_i} \frac{\partial M(l)}{\partial x_j}.
\end{equation}
If we assume now that $M(l)^{-1}$ has a pronounced peak at position $l^*$, as explained in the main text, then $M^{-4}(l)$ is even more peaked, and the sum is dominated by $l=l^*$:
\begin{equation}
\frac{\partial^2 d^2}{\partial x_i \partial x_j} \simeq \frac2{\pi^2 \tilde \kappa_0^2} \frac1{M^4(l^*)} \frac{\partial M(l^*)}{\partial x_i} \frac{\partial M(l^*)}{\partial x_j}.
\end{equation}

We now focus on the $(\hat{m},\hat{\sigma})$ subspace where the fit degeneracy arises. We recall that 
 \begin{equation}
M(l)=\hat{m}+ \frac{c_1^2 \hat{\sigma}}{l(l+1)-2+\hat{\sigma}} + 2\hat{J}l(l+1).
 \end{equation}
We get the Hessian matrix at the minimum of $d^2$:
 \begin{eqnarray}
{\rm Hess} (l^*) &=&  \frac2{\pi^2 \tilde \kappa_0^2} \frac1{M^4(l^*)} 
 \left(
\begin{array}{cc}
(\frac{\partial M}{\partial \hat m})^2 &  \frac{\partial M}{\partial \hat m} \frac{\partial M}{\partial \hat \sigma} \\ 
\\
\frac{\partial M}{\partial \hat \sigma} \frac{\partial M}{\partial \hat m} & (\frac{\partial M}{\partial \hat \sigma})^2
\end{array}
\right) \\
& = &
\left(
\begin{array}{cc}
1 & A(l^*) \\
A(l^*) & A^2(l^*) 
\end{array}
\right),
\end{eqnarray}
where
\begin{equation}
 A(l) =  \frac2{\pi^2 \tilde \kappa_0^2} \frac{c_1^2}{M^4(l)} \frac{l(l+1)-2}{[l(l+1) - 2 +\hat\sigma]^2}.
\end{equation}
 
\begin{figure}[t]
\begin{center}
\includegraphics[width=0.95\columnwidth]{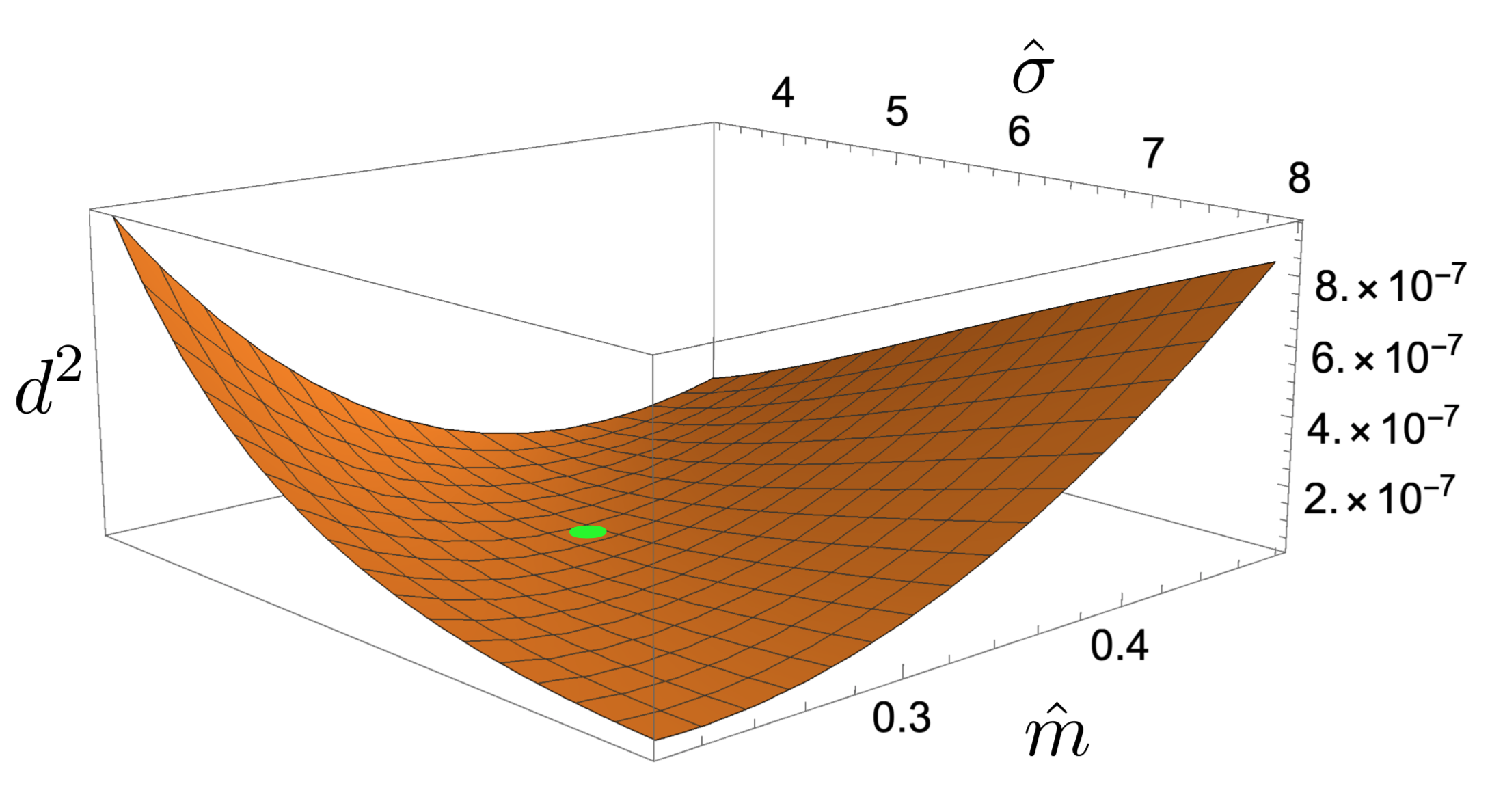} 
\end{center}
\caption{Valley of quasi-degenerate minima of the squared differences $d^2(\hat{J},\hat{m},\hat{\sigma})$ between the theoretical structure factor $S(l)$ and some measured values, plotted in the $(\hat{m},\hat{\sigma})$  subspace. The green dot represents the value found by the GOSA algorithm~\cite{Czaplicki, Goffe}. Here $c_1=1$, $\bar{\phi}=0.5$, $\tilde{J}_I^{-1}=4$ and $\tilde{\sigma}=300$.  
\label{valley}}
\end{figure}

This matrix has a trivial vanishing eigenvalue, which is the signature of a degenerate minimum, more precisely a valley of minima in the $(\hat{m},\hat{\sigma})$  subspace parallel to the corresponding eigenstate. Figure~\ref{valley}  illustrates this result. 
 
\section{Phase diagram frontier expressions}
\label{sec:front}

Here we write the frontier expressions for Eqs.~(\ref{eq:F1}) to (\ref{eq:F3}) with the dimensionless parameters and express $\tilde{J}_I^{-1}$ in function of $c_1$. Using \eq{hatJ}, \eq{hatm} and $\hat{\sigma}=\tilde{\sigma}/\tilde{\kappa_0}$, we get for \eq{eq:F1}
\beq
\tilde{J}_I^{-1}=\frac{4\sqrt{3}\tilde{\sigma}}{\tilde{\kappa}_0^2c_1^2}
\label{eq:F1b}
\eeq
and for \eq{eq:F2} we get
\beq
\tilde{J}_I^{-1}=\frac{\tilde{J}_{I,c}^{-1}-\frac{8\sqrt{3}}{\alpha_0 N}}{1+\frac{\tilde{\kappa}_0c_1^2}{\alpha_0 N}}
\label{eq:F2b}
\eeq
For \eq{eq:F3} it is more convenient to express $c_1$ in function of $\tilde{J}_I$:
\begin{eqnarray}
c_1&=&-\frac{\alpha_0 N}{4(\sqrt{3}\tilde{\sigma}\tilde{J}_I)^{1/2}}
\nonumber \\
&+ 
&\left(\frac{(\sqrt{3}\tilde{\sigma})^{1/2}}{\tilde{\kappa}_0}-2\sqrt{\frac{\sqrt{3}}{\tilde{\sigma}}}+ \frac{\alpha_0 N} {4\tilde{J}_{I,c}(\sqrt{3}\tilde{\sigma})^{1/2} } \right)
\sqrt{\tilde{J}_I}\qquad
\label{eq:F3b}
\end{eqnarray}

The coordinates of the triple point, corresponding to the intersection of these 3 curves of Eqs.~(\ref{eq:F1}, \ref{eq:F2}, \ref{eq:F3}) are
\begin{eqnarray}
c_1 &=& \left(\frac{\alpha_0 N \tilde{\sigma}}{\left[\alpha_0 N/(4\sqrt{3}\tilde{J}_{I,c}) - 2\right] \tilde{\kappa}_0^2 - \tilde{\sigma}\tilde{\kappa_0}} \right)^{1/2}\simeq1.23\qquad\\
\tilde{J}_I^{-1} &=& \tilde{J}_{I,c}^{-1} -4\sqrt{3} \frac{\tilde{\sigma}+2\tilde{\kappa}_0}{\alpha_0 N \tilde{\kappa}_0}\simeq3.46
\end{eqnarray}%
with $N=2562$, $\tilde{\sigma}=300$, $\alpha_0=1/\pi$ and $\tilde{J}_{I,c}^{-1}\simeq3.64$ on an infinite triangular lattice~\cite{Baxter}.

\section{Finite-size effects}
\label{sec:finite_size}

\begin{figure}[t]
\centering
\includegraphics[width=\columnwidth]{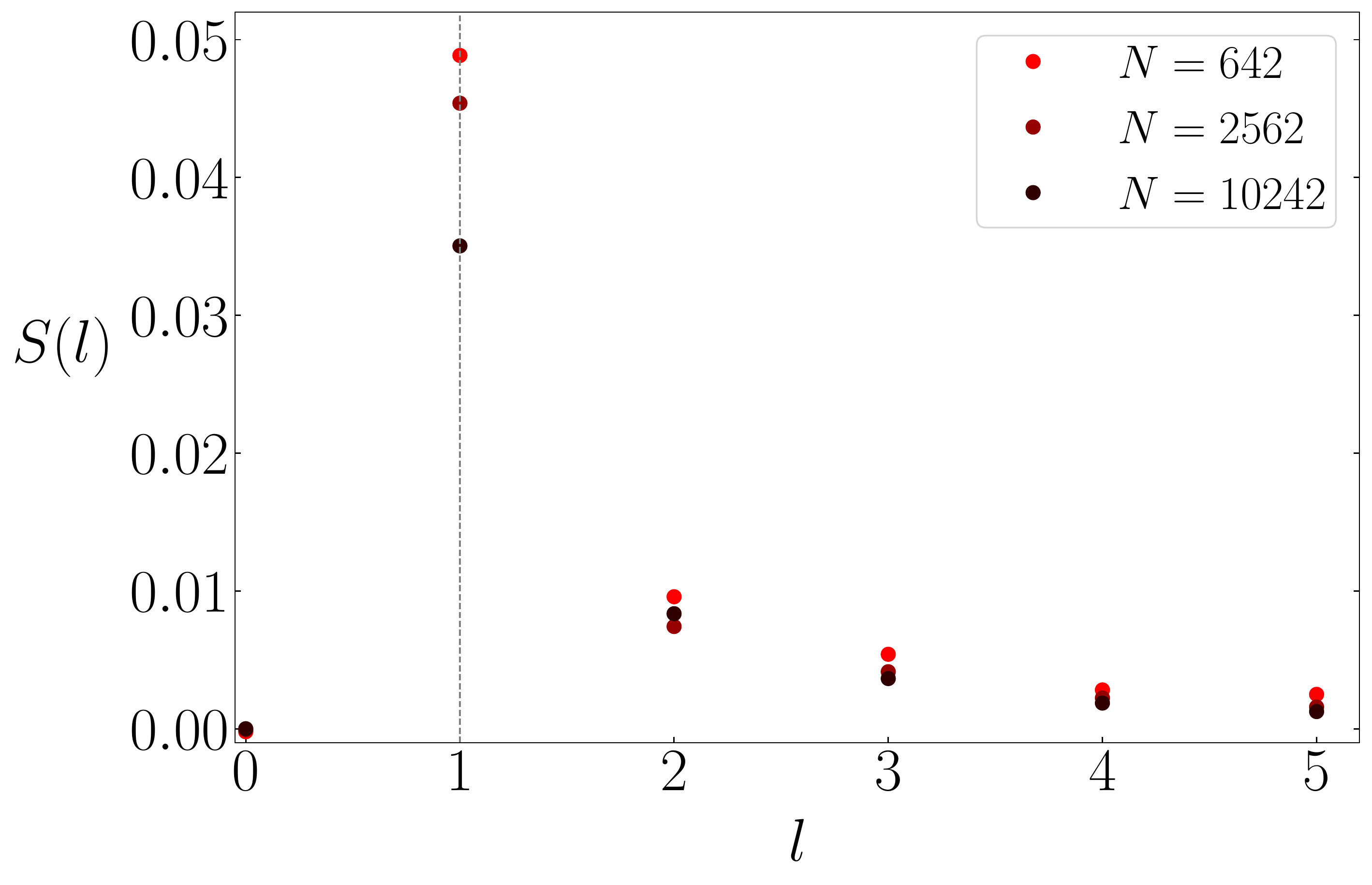} 
\caption{Structure factor for different system sizes $N=642$, 2562 and 10242, in the decoupled case (pure Ising model) at the critical point. }
\label{fig:syst-size}
\end{figure}

The theory developed in \refc{GG1} is valid for infinite-size systems, whereas we study finite-size ones. This has some consequences on physical observables.
The position of the second maximum in the structure factor indicates which mode is the most excited in the spatial species repartition $\phi(\theta,\varphi)$. This characterizes the patterns observed on the  vesicle. The $l=1$ mode is associated with the case where the species are totally separated, leading to one hemisphere rich in one species and one rich in the other (see the left-most vesicle in \fig{fig:effetC05}). Close to the critical point, large density fluctuations make possible the occurrence of very large clusters in an infinite system. In a finite-size system, this leads to an over-abundance of macro-clusters, as thoroughly studied, for example, in \refc{Toral}. This increases artificially the contribution of $l=1$ in the structure factor $S(l)$, as illustrated in \fig{fig:syst-size} where the decoupled case is studied for three different system sizes $N$. One observes that the bigger the system size, the lower the amplitude of the $l=1$ mode corresponding to a macrocluster.
This is the reason why we do not take this point into account when fitting $S(l)$.

\section*{Data availability statement}

The data that support the findings of this study are available from the corresponding author upon reasonable request.
\nocite{*}
\bibliography{Vesi1Leaflet}

\end{document}